\documentclass[12pt]{article} 
\usepackage{graphicx}
\usepackage{bm}

\begin{document}

\title{\bf Low-energy radiative-capture reactions 
within two-cluster coupled-channel description}  
\author{L.~Canton$^{1,2}$ and  L.~G.~Levchuk$^{2,3}$
}
\date{}
\maketitle
\begin{center}
{\em 
$^1$ Istituto Nazionale di Fisica Nucleare, 35131 Padova, via Marzolo, n. 8 Italy,\\
$^2$ Dipartimento di Fisica dell'Universit\`a, 35131 Padova, via Marzolo, n. 8 Italy,\\ 
$^3$ Kharkov Institute of Physics and Technology, 1, Akademicheskaya St., 
61108 Kharkov, Ukraine\\ 
}
\end{center}

\vspace{5mm}

\begin{abstract}
The formalism that describes radiative-capture reactions 
at low energies within an extended two-cluster potential model is 
presented. Construction of the operator of single-photon emission 
is based on a generalisation of the Siegert theorem 
with which the amplitude of the electromagnetic process is constructed
in an explicitly 
gauge-independent way. While the starting point for this construction 
is a microscopic (single-nucleon) current model, the resulting operator 
of low-energy photon emission by a two-cluster system is expressed in terms 
of macroscopic quantities for the clusters and does not depend directly
on their intrinsic coordinates and momenta.
The multichannel algebraic scattering (MCAS) approach has been used 
to construct the initial- and final-state wave functions. 
We present a general expression for the scattering wave function 
obtained from the MCAS $T$ matrix taking into account inelastic channels 
and Coulomb distortion. 
The developed formalism has been tested on the 
${^{3}{\rm He}}({\alpha},\gamma ){^{7}{\rm Be}}$ reaction cross section 
at astrophysical energies. The energy dependence of the evaluated cross section 
and $S$ factor agrees well with that extracted from measurement 
though the calculated quantities slightly overestimate data.
\end{abstract}




\maketitle

\newcommand{\ve}[1]{{\bm{#1}}}
\newcommand{\vegr}[1]{{\bm{#1}}}
\section{Introduction}
\label{sec1}
In spite of several decades of intensive experimental and theoretical 
studies of the radiative capture of nuclei at low-energies, reactions 
of this type remain a focus of research activities. 
One of the main reasons for this continued 
and growing interest in this kind of nuclear processes is their  
importance in understanding the Big Bang nucleo-synthesis and 
stellar evolution (particularly processes that occur in the interior 
of the Sun). Another aspect of astrophysical relevance that requires 
improved knowledge of the stellar fusion reactions,
is the solar neutrino flux problem. In particular,  
a recent assessment~\cite{Bahcall} 
of the current understanding of solar neutrino fluxes 
called for new measurements of 
the ${^{3}{\rm He}}({\alpha},\gamma ){^{7}{\rm Be}}$ reaction 
cross section at very low (solar) energies with an accuracy of better 
than $\pm$5\%. Of course, it is a serious challenge for an experiment
measuring cross sections at low energies 
to reach such a precision since the reaction yield rate is
extremely small due to the strong Coulomb repulsion.   

Some current challenges for nuclear astrophysics can be met with 
new experimental facilities such as the ones at 
the Laboratory for Underground Nuclear Astrophysics 
(LUNA)~\cite{LUNA} that is part of Italy's Gran Sasso National Laboratory 
(LNGS). The LUNA collaboration has reported on their first 
measurements~\cite{Bemmerer} 
of the ${^{3}{\rm He}}({\alpha},\gamma ){^{7}{\rm Be}}$ total 
cross section at very low (down to 127~keV) center-of-mass (c.m.) energies 
with a total uncertainty of 4\%. High-precision measurements of 
cross sections for other radiative-capture reactions of astrophysical 
relevance are under way~\cite{Prati}. However, some processes are, and
will remain the province of purely theoretical studies.
Nonetheless, the impressive progress achieved in recent times  
in the experimental studies of low-energy 
radiative-capture reactions is an additional argument for boosting research 
activities on the theoretical side as well. 

The photon-emission process relevant to stellar 
fusion reactions is known~\cite{ChristyDuck} 
to occur dominantly at a large (extra-nuclear) distance. 
In general, then, one may assume
that no intermediate ``compound'' nuclear system is formed.
If that is so,  
the process is defined as a direct-capture reaction. 
The ${^{3}{\rm He}}({\alpha},\gamma ){^{7}{\rm Be}}$ reaction, which provides 
termination of the proton-proton stellar chain through ${^{3}{\rm He}}$ 
burning, is a one case where
successful application of the direct-capture model 
as a description of the photon emission process has been made. 
One of the earliest detailed analysis of this reaction, and
of the mirror process ${^{3}{\rm H}}({\alpha},\gamma ){^{7}{\rm Li}}$, 
assuming a two-body direct-capture picture was made in Ref.~\cite{Tombrello}. 
In the calculations of the total cross section and of the astrophysical $S$ 
factor, the authors of that paper used 
a bound-state radial wave function taken in the asymptotic 
(Whittaker) form while a repulsive nuclear hard-core model that took   
into account Coulomb distortion was used to describe 
the initial (scattering) state. Such a choice for the initial-state model 
was favoured by the analysis of the then existing experimental data 
for elastic $\alpha$ scattering from the $A=3$ nucleus and the phase shifts
deduced therefrom. 
It was shown many years later~\cite{Kim81} that this repulsive effect 
can also result from inner oscillation in the overlap with the relative wave 
function of the two nuclei. Those later calculations were made
using both a two-body potential model and the orthogonality 
condition model (OCM)~\cite{Saito} to construct the wave function. 

The direct nature of many of stellar reactions suggests that
using a cluster model may be an appropriate approach for their description. 
In general, a great amount of experimental information on these 
reactions has been accumulated and extensive cluster-model 
studies of the stellar fusion processes have been made~\cite{Langanke94/86}. 
Methodologically, studies have ranged from the potential cluster 
models~\cite{Igamov07,Mohr93}) which consider 
the two clusters as point-like objects in a two-body potential, 
to microscopic approaches which include the resonating-group method 
(RGM)~\cite{WLKT83} as well as the generator coordinate method~\cite{Baye} 
and the OCM~\cite{Saito}, both of which are closely related to the RGM. 

In addition to the cluster models, the variational Monte Carlo technique 
has been elaborated~\cite{Marcucci} and applied in studies of 
few-nucleon systems. 
The dynamical input in this approach is given by microscopic 
(nucleon-nucleon) interactions.  Doing so provides a universality of 
description and reduces the number of model parameters. 
In an application of this approach~\cite{Nollett01},  
total cross sections
for the reactions ${^{3}{\rm He}}({\alpha},\gamma ){^{7}{\rm Be}}$ 
and ${^{3}{\rm H}}({\alpha},\gamma ){^{7}{\rm Li}}$ from calculations 
were found.
But this technique
could not reproduce adequately nuclear binding energies, 
and the states of ${^{7}{\rm Be}}$ were so under-bound as to be nonphysical. 
On the other hand, microscopic grounds to justify use of  
potential cluster models have been established within the so-called 
microscopic potential model~\cite{Langanke94/86}. Possible 
extensions of the two-cluster model space were also studied within 
that approach (see Ref.~\cite{Langanke00}). 

Nevertheless, many of essential dynamical features of low-energy 
radiative capture processes with light nuclei are still missing 
in the existing theoretical models. One of the important 
issues that has not been properly addressed so far 
is the role in the EM process of collective (rotational and vibrational) 
degrees of freedom of the colliding nuclei. 
This is a serious omission since
neglect of low-energy collective excitations and 
the corresponding resonances they produce can alter cross 
sections at low energies substantially. Thus the usual theoretical $S$-factor 
extrapolation from the quantities determined by measurement, to the energy 
region of astrophysical interest can be incorrect. 

Recently a multi-channel algebraic scattering (MCAS) approach~\cite{MCAS1} 
has been developed in which the interaction of a nucleon with a nucleus is 
described within the collective model~\cite{Tamura}. Extension of this 
approach to the treatment of the two-cluster systems 
has been made~\cite{MCAS2}. 
The MCAS model exploits the so-called orthogonalizing 
pseudo-potential (OPP) method~\cite{Kukulin} reformulated to take 
account of the influence of the Pauli exclusion principle in 
the collective model. 
This method is closely related to the OCM~\cite{Saito}, in which 
the physical two-body states are obtained by renormalization that
ensures their 
orthogonality to a certain number of unphysical bound states. The OPP 
method does not involve intrinsic degrees of freedom for the two 
fragments. Instead, an extra nonlocal contribution to the original two-body 
potential is constructed~\cite{MCAS1}. The MCAS approach solves the low energy
coupled equations for the interacting systems, and to use information gleaned
from such in capture reaction studies, the photon-emission operator 
has to be developed
within a two-cluster coupled-channel model that also takes into account 
cluster sizes and the explicit coupling to their intrinsic excitations. 

To construct such an operator, we start from the 
extension~\cite{LevSh93} of the Siegert theorem~\cite{Siegert}, expressing 
the electromagnetic (EM) current operator in terms of 
so-called generalised electric and magnetic moments of the nuclear system. 
Then, the amplitude of the EM transition is determined by  
matrix elements of that operator between nuclear states and 
the electric and magnetic field strengths. This explicitly gauge 
independent representation for the amplitude guarantees 
the fulfilment of the Siegert theorem  
for electric transitions in the long-wavelength limit. Decomposition 
of the EM operator into electric and magnetic parts is accomplished 
in a 3-dimensional form which 
assists in derivation of expressions for the low-energy 
operator of photon emission (or absorption) by a system of two extended 
clusters in terms of their macroscopic properties (the mean square radii, 
magnetic dipole moments and electric quadrupole moments) starting from 
the microscopic (single-particle) treatment of the total system. 

Herein we present expressions for the two-cluster 
EM operator and for calculation of observables for low-energy nuclear 
radiative capture instigated by this operator. 
The wave functions for the initial 
(scattering) and final (bound) states will be built within the MCAS scheme. 
In the next section, the manifestly gauge independent 
expression for 
the photon-emission amplitude based on the extension of the Siegert 
theorem is described. 
Sec.~\ref{sec3} contains derivation of the operator of 
single-photon emission by 
a system of two extended clusters, while the wave function 
construction according
to the MCAS method is detailed in Sec.~\ref{sec4}. 
Then, in Sec.~\ref{sec5}, results of application of 
this general formalism to describe 
the ${^{3}{\rm He}}({\alpha},\gamma ){^{7}{\rm Be}}$ reaction
are reported. 
Results for the cross 
section and the $S$ factor are discussed and compared with experimental 
data, and
notably that from the LUNA experiment~\cite{Bemmerer}. 
In the context of applications
of the MCAS approach, the reaction studied is considered as a simplified test 
calculation since it
does not yet involve inelastic channels. Our future plans are to use
this method to study
more complex reactions where inclusion of low-lying inelastic 
channels will be
an important aspect of the dynamics.
Concluding remarks are given in Sec.~\ref{sec6}.

\sloppy
\section{Photon-emission amplitude based on extension of the Siegert theorem} 
\label{sec2}

The amplitude of transition of a nuclear system from an initial state 
$\mid \ve{P}_i ; i \rangle$ to a state $\mid \ve{P}_f ; f \rangle$ 
with emission of a single photon with the energy 
(momentum) $E_{\gamma}$ ($\ve{k}$) and the polarization vector 
$\varepsilon_{\nu}(\ve{k}) \equiv (\varepsilon_{0}(\ve{k}),$ 
$\vegr{\varepsilon} (\ve{k}) )$ is given by 
\begin{equation}
{\cal T}_{if} = {(2\pi )}^{4} \delta (\ve{P}_i - \ve{P}_f -\ve{k} )\ 
\delta (E_i - E_f - E_{\gamma} )\ T_{if}, 
\label{Tauif}
\end{equation} 
\begin{equation}
T_{if} = -{1\over {(2\pi )^{3/2} \sqrt{2E_{\gamma}}}} 
\langle \ve{P}_f ; f \mid {\varepsilon}^{\nu} {J}_{\nu} (0) 
\mid \ve{P}_i ; i \rangle\ , 
\label{Tif1}
\end{equation} 
\begin{displaymath}
({\nu} = 0, 1, 2, 3)
\end{displaymath}
where $J_{\nu} (0)$ is the EM current operator at the space-time 
point $x=(t,\ve{x})=0$, $E_i$ and $\ve{P}_i$ ($E_f$ and $\ve{P}_f$) 
denote the total energy and momentum for the initial (final) state, and 
indices $i$ and $f$ refer to system ``intrinsic'' degrees of freedom. 

For the operator $J_{\nu} (\ve{x})$ and the Hamiltonian $H$ of the nuclear 
system, the continuity equation 
\begin{equation}
{\rm div}\ \ve{J}(\ve{x}) = i [ H, J_{0}(\ve{x}) ]\ ,
\label{CE}
\end{equation} 
has to hold. This gives rise to the gauge independence (GI) condition 
\begin{equation}
\ve{k} \langle \ve{P}_f ; f \mid \ve{J} (0) \mid \ve{P}_i ; i \rangle\ = 
(E_i - E_f) \langle \ve{P}_f ; f \mid  \rho (0) \mid \ve{P}_i ; i \rangle\ , 
\label{GI}
\end{equation} 
where ${\rho}(0)\equiv J_{0}(0)$, for the transition matrix element. 

In practical calculations, however, this condition is often violated 
because of shortcomings in the description of the nuclear dynamics. 
For instance, for a system consisting of $A_{\rm tot}$ 
nucleons , typically the EM current is given by the one-body 
contribution~\cite{BohrMottelson}, 
\begin{equation} 
\rho (0) = \rho^{[1]} \equiv \sum_{\xi =1}^{A_{\rm tot}} 
\widehat{e}_{\xi} \delta ({\ve{r}_{\xi}})\ , 
\label{ModelRho}
\end{equation} 
\begin{equation}
\ve{J} (0) = \ve{J}^{[1]} \equiv \sum_{\xi =1}^{A_{\rm tot}} 
{{\widehat{e}_{\xi}}\over {2m_{\xi}}} 
\{ {\ve{p}_{\xi}} , \delta ({\ve{r}_{\xi}})\}\ +\ 
i \sum_{\xi =1}^{A_{\rm tot}} e {{\widehat{\mu}_{\xi}}\over {m_{\xi}}} 
\left[ \vegr{s}_{\xi} \times \ve{p}_{\xi} , \delta ({\ve{r}_{\xi}})\right]\ , 
\label{ModelJ}
\end{equation} 
where $\widehat{e}_{\xi}=e$, $\widehat{\mu}_{\xi}=\mu_{p}$ 
($\widehat{e}_{\xi}=0$, $\widehat{\mu}_{\xi}=\mu_{n}$) for protons (neutrons), 
$e$ is the elementary charge, $\mu_{p}$ ($\mu_{n}$) denotes the proton 
(neutron) magnetic moment in nuclear magnetons, and $\ve{p}_{\xi}$, 
$\ve{r}_{\xi}$ and $\vegr{s}_{\xi}$ are the momentum, 
coordinate and spin operators for a nucleon number $\xi$. 
This current does not obey Eq.~(\ref{CE}), and one instead has 
\begin{equation}
[\ve{P}_{\rm tot}, \ve{J}(0) ] = [ K, \rho(0) ] 
= [ H, \rho(0) ] - [ V, \rho(0) ]\ , 
\label{NoCE}
\end{equation} 
\begin{displaymath}
\ve{P}_{\rm tot} = \sum_{\xi =1}^{A_{\rm tot}} \ve{p}_{\xi}\ ,
\end{displaymath}
where $K$ is the nuclear system kinetic-energy operator and $V$ denotes 
the nuclear potential. 
Evidently, the GI condition, Eq.~(\ref{GI}), is not met for the model
specified by Eqs.~(\ref{ModelRho}) and
(\ref{ModelJ}) if $V$ contains nonlocal (momentum-dependent) 
contributions. An interaction current has to be added 
to restore the gauge independence. 

Furthermore, it may happen that nuclear wave functions used in calculations are 
not eigenstates of $H$ corresponding to eigenvalues $E_i$ and $E_f$ for 
the initial and final state. That is the case in calculation of 
transition amplitudes for direct radiative capture when  
asymptotic radial wave functions are used for the initial and/or final states.
The same problem exists in any 
situation where, due to the increased complexity of the boundary
conditions, the scattering process is treated with a Hamiltonian 
that has been simplified with respect to that used for the nuclear
bound state.

In view of the difficulties, which are encountered in description of 
the nuclear system, 
the representation of the amplitude of an EM process discussed in 
Refs.~\cite{LevSh93,Foldy,FriarFall,Sheb89} can be useful. 
This representation expresses $T_{if}$ in 
a manifestly gauge-independent way in terms of 
the electric ($\ve{E}(\ve{k})$) 
and magnetic ($\ve{H}(\ve{k})$) field strengths, {\it viz.} 
\begin{equation}
T_{if} = \ve{E}(\ve{k})\ \ve{D}_{if}(\ve{k})\ 
+\ \ve{H}(\ve{k})\ \ve{M}_{if}(\ve{k})\ , 
\label{Tif2}
\end{equation} 
\begin{equation}
\ve{E}(\ve{k}) = -i [2(2\pi )^{3} E_{\gamma}]^{-1/2} \left( ( E_i - E_f ) 
\vegr{\varepsilon} (\ve{k}) - \ve{k} \varepsilon_0 (\ve{k})\right) \ ,  
\label{E}
\end{equation} 
\begin{equation}
\ve{H}(\ve{k}) = -i [2(2\pi )^{3} E_{\gamma}]^{-1/2} \ve{k} \times 
\vegr{\varepsilon} (\ve{k}) \ ,  
\label{H}
\end{equation} 
where $\ve{D}_{if} (\ve{k})$ and $\ve{M}_{if} (\ve{k})$ are matrix elements 
of the so-called generalised electric and magnetic dipole moments of 
the system determined by the original current $J_{\nu}$. 
This relationship provides an extension of the Siegert 
theorem~\cite{Siegert} and is an alternative to Eq.~(\ref{Tif1}). 
According to the representation, derivation of which relies 
upon Eq.~(\ref{CE}) and the diagonality of the total charge operator, 
we can redefine the current matrix element as
\begin{equation}
\langle \ve{P}_f ; f \mid  {\rho} (0) \mid \ve{P}_i ; i \rangle = 
-i \ve{k} \ve{D}_{if}(\ve{k})\ ,
\label{Curr20}
\end{equation} 
\begin{equation}
\langle \ve{P}_f ; f \mid  \ve{J}(0) \mid \ve{P}_i ; i \rangle = 
-i ( E_i - E_f )\ \ve{D}_{if}(\ve{k})\ - i \ve{M}_{if}(\ve{k})\times\ve{k} \ . 
\label{Curr2V}
\end{equation} 
This provides the automatic fulfilment of the GI condition, Eq.~(\ref{GI}).

Decomposition of the EM current operator into the electric and magnetic 
parts, which gives rise to Eq.~(\ref{Tif2}), has been
discussed in detail in Ref.~\cite{LevSh93}. 
Thus we give only a brief outline of 
the corresponding derivation and clarify the meaning of the quantities 
$\ve{D}_{if} (\ve{k})$ and $\ve{M}_{if} (\ve{k})$ by discussing their 
behaviour in the long-wavelength limit. In particular, we establish the link 
between 
the electric contribution in Eq.~(\ref{Tif2}) to the well-known Siegert 
result~\cite{Siegert}.  

So, we factor out the center-of-mass (c.m.) motion of the nuclear states, 
i.e. we use the separation 
\begin{equation}
\mid \ve{P}_i ; i \rangle\ =\ \mid \ve{P}_i ) \mid i \rangle\ ,\ \ \
\mid \ve{P}_i -\ve{k}; f \rangle\ =\ \mid \ve{P}_i -\ve{k} ) \mid f \rangle\ , 
\label{c.m.sep.}
\end{equation} 
which, of course, is always valid at non-relativistic energies. 
Then, the translation properties of the total momentum eigenstates allow us 
to write
\begin{equation}
 (\ve{P}_i -\ve{k} \mid =\  (\ve{P}_i \mid\ {\rm e}^{i\ve{k}\ve{R}_{\rm tot} }\ , 
\label{translation}
\end{equation} 
where $\ve{R}_{\rm tot}$ is the c.m. coordinate operator. 
The product of the exponent in 
Eq.~(\ref{translation}) and the vector $\vegr{\varepsilon} (\ve{k})$ can be 
decomposed into the ``electric'' and ``magnetic'' parts using 
the identity (cf. Ref.~\cite{Foldy})
\begin{equation}
\vegr{\varepsilon} {\rm e}^{i\ve{k} \ve{R}_{\rm tot}} = 
\int_0^1 \{ [ {\ve{P}_{\rm tot}} , 
i\vegr{\varepsilon} {\ve{R}_{\rm tot}} {\rm e}^{i\lambda\ve{k}{\ve{R}_{\rm tot}}} ]\ +\ 
i\lambda {\ve{R}_{\rm tot}}\times [ \vegr{\varepsilon} \times \ve{k} ] 
{\rm e}^{i\lambda \ve{k}{\ve{R}_{\rm tot}}} \}\ {\rm d} \lambda\ ,  
\label{trick1}
\end{equation} 
where the components of vectors $\ve{P}_{\rm tot}$ and $\ve{R}_{\rm tot}$ obey 
the usual commutation rule 
$$
[{P}_{{\rm tot},j} , {R}_{{\rm tot},l} ] = -i \delta_{jl}\ , \ \ \ \ 
(j,l =1,2,3)\ .
$$   
Analogously, 
\begin{equation}
\varepsilon_{0} {\rm e}^{i\ve{k}\ve{R}_{\rm tot}} = 
\varepsilon_{0} \left( 1 + {i\ve{k}\ve{R}_{\rm tot}} 
\int_0^1 {\rm e}^{i\lambda\ve{k}\ve{R}_{\rm tot}} \ {\rm d} \lambda \right)\ .
\label{trick2}
\end{equation} 
Taking into account the continuity equation, Eq.~(\ref{CE}), 
and as the total charge 
operator is diagonal, from Eqs.~(\ref{Tif1}), (\ref{trick1}) 
and (\ref{trick2}) we deduce Eq.~(\ref{Tif2}) with 
\begin{equation}
\ve{D}_{if}(\ve{k}) = (2{\pi})^{-3} 
\langle f \mid \ve{D} (\ve{k}) \mid i \rangle\ ,
\label{Dif}
\end{equation} 
\begin{equation}
\ve{M}_{if}(\ve{k}) = (2{\pi})^{-3} 
\langle f \mid \ve{M} (\ve{k}) \mid i \rangle\ . 
\label{Mif}
\end{equation} 
The operators 
\begin{equation}
\ve{D} (\ve{k}) = - {{(2{\pi})^{3}}\over {E_i - E_f}}\int_0^1 
(\ve{P}_i - \lambda\ve{k} \mid \ve{R}_{\rm tot} \left[ \rho (0) , H\right]\ \mid 
\ve{P}_i )  \ {\rm d} \lambda\ ,
\label{D}
\end{equation} 
\begin{equation}
\ve{M} (\ve{k}) = - (2{\pi})^{3} \int_0^1 
(\ve{P}_i - \lambda\ve{k} \mid \ve{R}_{\rm tot} \times \ve{J} (0) \ \mid 
\ve{P}_i )  \ \lambda{\rm d} \lambda 
\label{M}
\end{equation} 
act in the space of the intrinsic variables of the nuclear states. 
It should be stressed that Eq.~(\ref{Tif2}) is quite 
general and can be used to describe various EM processes 
with nuclei at low and intermediate energies. In case of the inverse reaction 
(absorption of the photon with momentum $\ve{k}$), one has to replace 
$\lambda\ve{k}$ by $-\lambda\ve{k}$ in Eqs.~(\ref{D}) and (\ref{M}). 
The results given by Eqs.~(\ref{Tif1}) and (\ref{Tif2}) are identical 
if the GI condition of Eq.~(\ref{GI}) holds, but can be different otherwise. 
We choose the 
representation of the amplitude given by Eq.~(\ref{Tif2}) since it provides 
automatically the GI of calculations and helps to reproduce predictions of 
the low-energy theorems based on the gauge invariance. 
Furthermore, 
as will be shown in Sec.~\ref{sec3}, this form of the amplitude 
is a convenient starting point for construction of effective EM 
interactions for a system of extended nuclear objects (clusters) with 
an account for their internal structure. 

The representation through Eq.~(\ref{Tif2}) was successfully employed in our 
studies~\cite{EPJA04andLevSh99} of pion photo- and electroproduction 
off nucleons and light nuclei. The current paper addresses construction 
of EM operators for applications at very low energies (such 
as nuclear radiative-capture reactions at energies of astrophysical interest). 

Operators defined by Eqs.~(\ref{D}) and (\ref{M}) act in the space of states 
characterised 
by intrinsic coordinates 
\begin{equation}
\ve{r'}_{\xi} = \ve{r}_{\xi} - \ve{R}_{\rm tot}  
\label{r'}
\end{equation} 
and momenta
\begin{equation}
\ve{p'}_{\xi} = \ve{p}_{\xi} - {{m_{\xi}}\over {M}}\ve{P}_{\rm tot} \ ,  
\label{p'}
\end{equation} 
\begin{displaymath}
M = \sum_{\xi =1}^{A_{\rm tot}} m_{\xi}\ , 
\end{displaymath}
with commutation relations 
\begin{equation}
[ \ve{r'}_{\xi , j} , \ve{p'}_{\kappa , l} ] =   
i (\delta_{\xi\kappa} - {{m_{\kappa}}\over {M}})\ \delta_{jl} 
\label{[r',p']}
\end{equation} 
\begin{displaymath}
(\xi ,\kappa =1,2,...,A_{\rm tot}\ ;\ \ \ \ j,l = 1,2,3)\ .
\end{displaymath}
To clarify the meaning of the quantities specified in Eqs.~~(\ref{D}) and (\ref{M}), 
consider the current model as defined by Eqs. ~(\ref{ModelRho}) and (\ref{ModelJ}). 
After substitution of Eqs.~(\ref{ModelRho}) and (\ref{ModelJ}) into Eqs.~(\ref{D}) 
and (\ref{M}) and taking into account relations given in Eqs.~(\ref{r'})~-- 
(\ref{[r',p']}), 
we obtain 
\begin{eqnarray}
& & \ve{D} (\ve{k})  
= {{1}\over {E_i - E_f}}\int_0^1 {\rm d} \lambda \{
\left[ \ve{D}_{\rm int} (\lambda\ve{k}) , H_{\rm int} \right] 
\nonumber \\ 
& &+ {{\lambda\ve{k} (2\ve{P}_{i} - \lambda\ve{k} )}\over {2M}} 
\ve{D}_{\rm int} (\lambda\ve{k}) + 
i {{\ve{P}_{i} - \lambda\ve{k} }\over {M}} \rho_{\rm int} (\lambda\ve{k}) 
\} \ ,
\label{2pi3D}
\end{eqnarray} 
\begin{equation}
\rho_{\rm int} (\lambda\ve{k}) = 
\sum_{\xi =1}^{A_{\rm tot}} \widehat{e}_{\xi} {\rm e}^{-i\lambda\ve{k}
\ve{r'}_{\xi} } \ , 
\label{rhoint}
\end{equation} 
\begin{equation}
\ve{D}_{\rm int} (\lambda\ve{k}) = 
\sum_{\xi =1}^{A_{\rm tot}} \widehat{e}_{\xi} \ve{r'}_{\xi} 
{\rm e}^{-i\lambda\ve{k}\ve{r'}_{\xi} } 
\label{Dint}
\end{equation} 
for the electric generalised dipole operator and 
\begin{equation}
\ve{M} (\ve{k}) = 
\ve{M}^{\rm orb} (\ve{k}) + \ve{M}^{\rm spin} (\ve{k}) \ ,
\label{2pi3M}
\end{equation} 
\begin{equation}
\ve{M}^{\rm orb} (\ve{k}) = \int_0^1 \lambda {\rm d} \lambda \{
\ve{M}_{\rm int}^{\rm orb} (\lambda\ve{k}) - 
{{2\ve{P}_{i} - \lambda\ve{k}}\over {2M}}\times \ve{D}_{\rm int} (\lambda\ve{k})
\} \ ,
\label{Morb}
\end{equation} 
\begin{equation}
\ve{M}^{\rm spin} (\ve{k}) = \sum_{\xi =1}^{A_{\rm tot}} e 
{{\widehat{\mu}_{\xi}}\over {m_{\xi}}} \ve{s}_{\xi} 
{\rm e}^{-i\ve{k}\ve{r'}_{\xi} } + 
\ve{k} \int_0^1 \lambda^{2} {\rm d} \lambda \left(
\sum_{\xi =1}^{A_{\rm tot}} e 
{{\widehat{\mu}_{\xi}}\over {m_{\xi}}} (\ve{r'}_{\xi} \ve{s}_{\xi} ) 
{\rm e}^{-i\lambda\ve{k}\ve{r'}_{\xi} } 
\right) \ ,
\label{Mspin}
\end{equation} 
\begin{equation}
\ve{M}_{\rm int}^{\rm orb} (\lambda\ve{k}) = \sum_{\xi =1}^{A_{\rm tot}} 
{\widehat{e}_{\xi}\over {2m_{\xi}}} 
\{
\ve{l'}_{\xi} , {\rm e}^{-i\lambda\ve{k}\ve{r'}_{\xi} } 
\} \ ,
\label{Morb/int}
\end{equation} 
\begin{equation}
\ve{l'}_{\xi} \equiv \ve{r'}_{\xi} \times \ve{p'}_{\xi} 
\label{l'}
\end{equation} 
for the magnetic one. Obviously, the second term in Eq.~(\ref{Mspin}) 
does not contribute to the amplitude in Eq.~(\ref{Tif2}). Moreover, 
for a $\ve{k}$-congruent frame ($\ve{P}_{i}\times\ve{k}=0$) which will be 
implied below, contributions 
from the second term of Eq.~(\ref{2pi3D}) and from the integrand in 
Eq.~(\ref{Morb}) cancel when emission of a real photon is considered. 
In that case one usually uses the Coulomb gauge,  
\begin{equation}
\varepsilon_{0}(\ve{k}) = 0\ , \ \ \ \ 
\ve{k}\vegr{\varepsilon}(\ve{k}) = 0\ ,
\label{CoulombGauge}
\end{equation} 
for which the last term in Eq.~(\ref{2pi3D}) also vanishes. 
Partial cancellation between the two terms of Eq.~(\ref{Tif2}) reflects 
the fact that, in general, decomposition of an EM transition amplitude 
into electric and magnetic parts is not unique 
(see, e.g., discussion 
in Ref.~\cite{FriarFall}). 

Thus, as a result, we have
\begin{equation}
\ve{D} (\ve{k})  
= {{E_i^{\rm int} - E_f^{\rm int}}\over {E_i - E_f}}
\int_0^1 \ve{D}_{\rm int} (\lambda\ve{k})\ {\rm d} \lambda\ ,
\label{GEDM}
\end{equation} 
\begin{equation}
\ve{M} (\ve{k}) = 
\int_0^1 \ve{M}_{\rm int}^{\rm orb} (\lambda\ve{k})\ \lambda {\rm d} \lambda + 
\sum_{\xi =1}^{A_{\rm tot}} e {{\widehat{\mu}_{\xi}}\over {m_{\xi}}} 
\vegr{s}_{\xi} {\rm e}^{-i\ve{k}\ve{r'}_{\xi} }\ , 
\label{GMDM}
\end{equation} 
where $E_i^{\rm int}$ ($E_f^{\rm int}$) is the intrinsic energy of the initial 
(final) nuclear system. Calculation of the amplitude of interest is reduced 
to calculation of the matrix elements of operators $\ve{D}_{\rm int}$, 
$\ve{M}_{\rm int}^{\rm orb}$ and $\ve{M}^{\rm spin}$ between the intrinsic 
states of the nuclear system. 
From Eq.~(\ref{GEDM}) 
the classical Siegert result~\cite{Siegert} for the electric transitions 
in the long-wavelength limit $\ve{k}\rightarrow 0$ is readily deduced. 
In this limit, the operator $\ve{M} (\ve{k})$ 
becomes that of the total magnetic moment for 
the system.  Unlike the common procedure~\cite{BohrMottelson,BrussGlaud} 
of obtaining the electric and magnetic multipole operators, 
Eq.~(\ref{Tif2}) with 
quantities $\ve{D}_{if}$ and $\ve{M}_{if}$ determined by 
Eqs.~(\ref{GEDM}) and (\ref{GMDM}) decomposes the amplitude into the 
electric and magnetic parts without using 
partial-wave expansion, and no approximations valid only for low photon 
energies are made. 
The operators given in Eqs.~(\ref{GEDM}) and (\ref{GMDM}) are 
three-dimensional involving an integration over $\lambda$ 
that can be easily carried out. 
In the next section we show that this representation 
in terms of generalised electric and magnetic dipole operators,
is of particular use in derivation of
photon-emission operators for cluster-type nuclear structures. 
In fact, with Eqs.~(\ref{GEDM}) and (\ref{GMDM}), 
general expressions for the operator of photon 
emission by a system of two (or more) finite-size composite 
objects can be derived while taking their internal cluster structure
into account. 

\sloppy
\section{Operator of single-photon emission by a two-cluster system} 
\label{sec3}

Assume now that the nuclear system consists of two 
subsystems (clusters) A ($A$ nucleons) and B ($B$=$A_{\rm tot}-A$ nucleons) 
and is described by a wave function determined by subsystem intrinsic 
states $\mid\phi^{\rm A}\rangle$ and $\mid\phi^{\rm B}\rangle$ and a vector 
$\mid\Psi_{\rm AB}\rangle$ corresponding to the di-cluster relative motion. 
We can introduce for such a system a new set of coordinates and momenta 
as follows 
\begin{equation}
\ve{r'}_{\alpha} = 
\ve{r}_{\alpha}^{\rm A} + {{M_{\rm B}}\over{M}}\ve{R}\ , \ \ \ 
\ve{r'}_{\beta} = 
\ve{r}_{\beta}^{\rm B} - {{M_{\rm A}}\over{M}}\ve{R}\ , 
\label{rArB}
\end{equation} 
\begin{equation}
\ve{p'}_{\alpha} =
\ve{p}_{\alpha}^{\rm A} +{{m_{\alpha}}\over{M_{\rm A}}}\ve{P}\ , 
\ \ \ 
\ve{p'}_{\beta} = 
\ve{p}_{\beta}^{\rm B} - {{m_{\beta}}\over{M_{\rm B}}}\ve{P}\ , 
\label{pApB}
\end{equation} 
\begin{equation}
\ve{R} = \ve{R}_{\rm A} - \ve{R}_{\rm B}\ , 
\ \ \ 
\ve{P} = 
{{M_{\rm B} \ve{P}_{\rm A} - M_{\rm A} \ve{P}_{\rm B}}\over{M}}\ , 
\label{RAB_PAB_def}
\end{equation} 
\begin{displaymath}
( \alpha = 1,2,...,A ;\ \ \ \beta = 1,2,...,B ) 
\end{displaymath}
where $\ve{R}_{\rm A,B}$ ($\ve{P}_{\rm A,B}$) and $M_{\rm A,B}$ are 
the coordinate (momentum) and mass of the subsystem A, B; 
and $\ve{r}_{\alpha}^{\rm A}$, $\ve{r}_{\beta}^{\rm B}$ 
($\ve{p}_{\alpha}^{\rm A}$, $\ve{p}_{\beta}^{\rm B}$) refers to 
the ``intrinsic'' coordinate (momentum) of a nucleon, 
which belongs to cluster A, B.  

For the two-cluster relative coordinate $\ve{R}$ and momentum 
$\ve{P}$ and the subsystem intrinsic operators, the following 
commutation relations hold 
\begin{eqnarray}
& &[ R_{j} , P_{l} ] =  i \delta_{jl}\ ,\ \ \ 
[ r_{\xi , j}^{\rm A,B} , p_{\kappa , l}^{\rm A,B} ] =   
i (\delta_{\xi\kappa} - {{m_{\kappa}}\over {M_{\rm A,B}}})\ \delta_{jl}\ ,
\nonumber\\
& &
[ r_{\xi , j}^{\rm A,B} , p_{\kappa , l}^{\rm B,A} ] =   
[ r_{\xi , j}^{\rm A,B} , P_{l} ] =   
[ R_{j} , p_{\kappa , l}^{\rm A,B} ] = 0 \ ,  
\label{NewComm}
\end{eqnarray} 
\begin{displaymath}
(\xi ,\kappa =1,2,...,A\ [B]\ {\rm for\ cluster\ A\ [B]};\ \ \ \ 
j,l = 1,2,3) \ .
\end{displaymath}
They can be obtained straightforwardly from  
Eqs.~(\ref{rArB})~-- (\ref{RAB_PAB_def}) with the help of Eq.~(\ref{[r',p']}) 
and manifest the reciprocal independence of the intrinsic coordinates and 
momenta for the two different subsystems from each other. 

Then, substitution of Eqs.~(\ref{rArB}) and (\ref{pApB}) into 
Eq.~(\ref{Dint}) gives 
\begin{equation}
\ve{D}_{\rm int} (\lambda\ve{k}) = 
\ve{d}_{\rm A} (\lambda\ve{k})\ 
{\rm e}^{-i\lambda{{M_{\rm B}}\over{M}}\ve{k}\ve{R}} + 
\ve{d}_{\rm B} (\lambda\ve{k})\ 
{\rm e}^{i\lambda{{M_{\rm A}}\over{M}}\ve{k}\ve{R}} + 
\ve{D}_{\rm AB} (\lambda\ve{k}) \ , 
\label{DintClust}
\end{equation} 
\begin{equation}
\ve{d}_{\rm A,B} (\lambda\ve{k}) =  
\sum_{\xi =1}^{A,B} \widehat{e}_{\xi} \ve{r}_{\xi}^{\rm A,B} 
{\rm e}^{-i\lambda\ve{k}\ve{r}_{\xi}^{\rm A,B}} \ , 
\label{dA,B}
\end{equation} 
\begin{equation}
\ve{D}_{\rm AB} (\lambda\ve{k}) = 
{{M_{\rm B}}\over{M}}\rho_{\rm A}(\lambda\ve{k})\ \ve{R}\ 
{\rm e}^{-i\lambda{{M_{\rm B}}\over{M}}\ve{k}\ve{R}} - 
{{M_{\rm A}}\over{M}}\rho_{\rm B}(\lambda\ve{k})\ \ve{R}\ 
{\rm e}^{i\lambda{{M_{\rm A}}\over{M}}\ve{k}\ve{R}} \ , 
\label{DAB}
\end{equation} 
\begin{equation}
\rho_{\rm A,B} (\lambda\ve{k}) =  
\sum_{\xi =1}^{A,B} \widehat{e}_{\xi} 
{\rm e}^{-i\lambda\ve{k}\ve{r}_{\xi}^{\rm A,B}} \ .  
\label{rhoA,B}  
\end{equation} 
It should be stressed that Eq.~(\ref{GEDM}) for the generalised 
electric dipole operator, with $\ve{D}_{\rm int} (\lambda\ve{k})$ determined  
by Eqs.~(\ref{DintClust})~-- (\ref{rhoA,B}), is valid to exactly 
the same extent as is the one-body current of 
Eqs.~(\ref{ModelRho}) and (\ref{ModelJ}). 
Eq.~(\ref{DintClust}) provides a convenient starting 
point to construct electric transition operators of different form
depending on a particular situation. 
Consider, for instance, low-energy photon emission 
by a system of two charged clusters with substantially different 
sizes. In addition, due to a strong Coulomb 
repulsion, on average the emission (absorption) process occurs 
at a very large separation distance, considerably greater than 
the characteristic size of the bigger cluster. 
Then, one can make a low-energy expansion of  
Eq.~(\ref{DintClust}) with respect to specific dynamical variables,
that is independent of the remaining variables.
In other words, it is possible
to obtain evaluations at different perturbative orders 
with respect to the three separated expansion parameters 
$\ve{k}\ve{r}_{\alpha}^{\rm A}$, $\ve{k}\ve{r}_{\beta}^{\rm B}$ and 
$\ve{k}\ve{R}$.  

In general, the quantity $\ve{d}_{\rm A,B} (\lambda\ve{k})$ 
in Eq.~(\ref{DintClust}) gives rise 
to a generalised electric dipole operator for the subsystem A, B  
and contains all multipole contributions responsible for the electric 
transitions in this subsystem. Analogously, the operator 
$\ve{D}_{\rm AB} (\lambda\ve{k})$ provides (for all orders in $\ve{k}$) 
the electric transition contribution due to the relative motion 
of the two clusters. 
But, as follows from Eq.~(\ref{DintClust}), 
the separation between the ``intrinsic'' (for the subsystems A and B) 
transitions and the ones due to the relative di-cluster motion 
is not complete. The ``intrinsic'' electric operators 
$\ve{d}_{\rm A} (\lambda\ve{k})$ and $\ve{d}_{\rm B} (\lambda\ve{k})$ 
in this formula are coupled to the exponential factors dependent on 
the two-cluster relative coordinate, while the quantity 
$\ve{D}_{\rm AB} (\lambda\ve{k})$ includes cluster form-factor 
operators $\rho_{\rm A} (\lambda\ve{k})$ and $\rho_{\rm B} (\lambda\ve{k})$. 

However, such a separation is exact in the long-wavelength limit 
in which, from Eqs.~(\ref{GEDM}) and (\ref{DintClust}), one gets 
\begin{equation}
\ve{D} (0)  = 
\ve{d}^{\rm A} \ + \ve{d}^{\rm B}\  + \ve{D}^{\rm AB} \ , 
\label{D(0)}
\end{equation} 
\begin{equation}
\ve{D}^{\rm AB}    = 
e C_{E1} \ve{R} \ , 
\label{DAB(0)}
\end{equation} 
\begin{equation}
C_{E1} =  
{{M_{\rm B} Z_{\rm A} - M_{\rm A} Z_{\rm B}}\over{M}}\ , 
\label{C_E1}
\end{equation} 
where $\ve{d}^{\rm A,B} \equiv \ve{d}_{\rm A,B} (0)$ is
the usual electric dipole operator of the system A, B~\cite{BohrMottelson,BrussGlaud}; 
$Z_{\rm A}$ ($Z_{\rm B}$) 
is the component A (B) charge, and $\ve{D}^{\rm AB}$ is 
the electric dipole operator for two point-like clusters.  
This proves that the constructed electric 
operator fulfils the Siegert theorem~\cite{Siegert} 
at $\ve{k}\rightarrow 0$. 
Note also that there is no recoil in this limit. Therefore 
the energy factor in Eq.~(\ref{GEDM}) is equal to 1. 

If the states $\mid\chi_{\rm A}\rangle$ and 
$\mid\chi_{\rm B}\rangle$ possess a definite parity, and 
the intrinsic parities of the clusters do not change in the
EM transitions, by neglecting terms of order 
$O({k^2})$ in Eq.~(\ref{DintClust}) we obtain
\begin{equation}
\ve{D}(\ve{k}) = {{E_i^{\rm int} - E_f^{\rm int}}\over {E_i - E_f}} 
\left( 
\ve{D}^{E1} - {{i}\over{6}} \ve{D}^{E2} (\ve{k})
\right) \ , 
\label{D(1)}
\end{equation} 
\begin{equation}
\ve{D}^{E1} \equiv \ve{D}^{\rm AB} 
\ , 
\label{D_E1}
\end{equation} 
\begin{equation}
{D}_{j}^{E2} (\ve{k}) = \sum_l
{k}_{l} \left( 
t_{jl}^{\rm A} + t_{jl}^{\rm B}  + e C_{E2} T_{jl}^{\rm AB} 
\right) \ , 
\label{D_E2}
\end{equation} 
\begin{equation}
t_{jl}^{\rm A,B} = \sum_{\xi =1}^{A,B} \widehat{e}_{\xi} 
\left( 3\ r_{\xi ,j}^{\rm A,B}\ r_{\xi ,l}^{\rm A,B} - 
\delta_{jl} \left[ r_{\xi}^{\rm A,B}\right] ^{2}
\right) \ ,
\label{tA,B}
\end{equation} 
\begin{equation}
T_{jl}^{\rm AB} = 
3 R_{j} R_{l} - \delta_{jl}  R^{2} 
 \ ,
\label{TAB}
\end{equation} 
\begin{equation}
C_{E2} = {{M_{\rm B}^2 Z_{\rm A} + M_{\rm A}^2 Z_{\rm B}}\over{M^2}} \ ,
\label{CE2}
\end{equation} 
\begin{displaymath}
( j,l = 1, 2, 3 )\ .
\end{displaymath}
Here, we have taken into  
account the gauge condition, Eq.~(\ref{CoulombGauge}). The operator in 
the parentheses of Eq.~(\ref{D_E2}) is the  
electric quadrupole tensor operator for the two-cluster system. 

Of course, one can proceed with analysis of Eq.~(\ref{DintClust}) 
and obtain higher-order contributions to the electric operator defined by 
Eq.~(\ref{GEDM}). 
In particular, a correction 
to the two-cluster electric dipole operator $\ve{D}^{\rm AB}$ emerges due to 
the finite cluster sizes. This correction can be of great 
importance when the radiative capture occurs with 
different symmetric nuclei, namely, 
those having equal numbers of protons and neutrons. For those cases, 
the quantity given by Eq.~(\ref{C_E1}) is vanishingly small. 
Examples are the radiative capture of ${^{4}{\rm He}}$ by 
${^{12}{\rm C}}$, by ${^{16}{\rm O}}$, or by some heavier nuclei. 
That capture process 
is very important in assessing  stellar evolution and nucleo-synthesis. 
For simplicity, assume that the reaction is analysed within a potential 
cluster model with no coupling to intrinsic excitations of the 
subsystems A and B ({\it i.e.}, they are considered as ``frozen''), 
and that the cluster electric quadrupole moments are zero. Then, for sufficiently 
small photon energy ($\ve{k}\ve{r}_{\xi}^{\rm A,B}\ll 1$),  
from Eq.~(\ref{DintClust}) the correction 
\begin{equation}
C_{E1} \rightarrow C_{E1} (k) \equiv C_{E1} + k^2 \Delta C_{E1} \ ,
\label{CE1Corr}
\end{equation}
\begin{equation}
\Delta C_{E1} = - {{1}\over{18}} 
{{M_{\rm B} Z_{\rm A} \langle {r}_{\rm A}^2 \rangle - 
M_{\rm A} Z_{\rm B} \langle {r}_{\rm B}^2 \rangle }\over{M_{\rm A}+M_{\rm B}}} \ ,
\label{DeltaC_E1}
\end{equation}
is found. Therein
$\langle {r}_{\rm A,B}^2 \rangle$ is the mean square charge radius 
of the cluster A, B. 
However, the corresponding 
contribution to the operator $\ve{D}(\ve{k})$ has to be taken into 
account together with other terms in Eq.~(\ref{GEDM}) of the same 
order in~$\ve{k}$. 

The same reasoning can be also applied to the magnetic 
operator given in Eq.~(\ref{GMDM}). 
Substituting Eqs.~(\ref{rArB}) and (\ref{pApB}) 
into Eq.~(\ref{Morb/int}) gives
\begin{eqnarray}
&&
\ve{M}_{\rm int}^{\rm orb} (\lambda\ve{k}) = 
\ve{m}_{\rm A}^{\rm orb} (\lambda\ve{k})\ 
{\rm e}^{-i\lambda{{M_{\rm B}}\over{M}}\ve{k}\ve{R}} + 
\ve{m}_{\rm B}^{\rm orb} (\lambda\ve{k})\ 
{\rm e}^{i\lambda{{M_{\rm A}}\over{M}}\ve{k}\ve{R}} + 
\ve{M}_{\rm AB}^{\rm orb} (\lambda\ve{k}) 
\nonumber \\
&&
+ 
{{M_{\rm B}}\over{2M}} \{ \ve{d}_{\rm A} (\lambda\ve{k})\times \ve{V},\  
{\rm e}^{-i\lambda{{M_{\rm B}}\over{M}}\ve{k}\ve{R}}\} - 
{{M_{\rm A}}\over{2M}}\{ \ve{d}_{\rm B} (\lambda\ve{k})\times \ve{V},\  
{\rm e}^{i\lambda{{M_{\rm A}}\over{M}}\ve{k}\ve{R}}\} 
\nonumber \\
&&
+ {{M_{\rm B}}\over{M}} 
\left[ \ve{R} \times \ve{v}_{\rm A}(\lambda\ve{k})\right] \  
{\rm e}^{-i\lambda{{M_{\rm B}}\over{M}}\ve{k}\ve{R}} - 
{{M_{\rm A}}\over{M}} 
\left[ \ve{R} \times \ve{v}_{\rm B}(\lambda\ve{k})\right] \  
{\rm e}^{i\lambda{{M_{\rm A}}\over{M}}\ve{k}\ve{R}}  
\ , 
\label{MintClust}
\end{eqnarray} 
\begin{equation}
\ve{m}_{\rm A,B}^{\rm orb} (\lambda\ve{k}) = \sum_{\xi =1}^{A,B} 
{\widehat{e}_{\xi}\over {2m_{\xi}}} 
\{
\ve{l}_{\xi}^{\rm A,B} , {\rm e}^{-i\lambda\ve{k}\ve{r}_{\xi}^{\rm A,B} } 
\} \ ,
\label{m_orb_A,B}
\end{equation} 
\begin{eqnarray}
&&
\ve{M}_{\rm AB}^{\rm orb} (\lambda\ve{k}) = 
{{1}\over{2M}} {{ M_{\rm B}}\over{M_{\rm A}}} {\rho}_{\rm A}(\lambda\ve{k}) 
\ \{ \ve{L},  
{\rm e}^{-i\lambda{{M_{\rm B}}\over{M}}\ve{k}\ve{R}}\} 
+
{{1}\over{2M}} {{M_{\rm A}}\over{M_{\rm B}}} {\rho}_{\rm B}(\lambda\ve{k}) 
\ \{ \ve{L},   
{\rm e}^{i\lambda{{M_{\rm A}}\over{M}}\ve{k}\ve{R}}\} 
\ ,
\label{M_orb_AB}
\end{eqnarray} 
\begin{equation}
\ve{v}_{\rm A,B} (\lambda\ve{k}) = {{1}\over{2}} \sum_{\xi =1}^{A,B} 
\widehat{e}_{\xi} \{ \ve{v}_{\xi}^{\rm A,B} ,
{\rm e}^{-i\lambda\ve{k}\ve{r}_{\xi}^{\rm A,B} } 
\} \ ,
\label{v_A,B}
\end{equation} 
\begin{equation}
\ve{v}_{\xi}^{\rm A,B} =  
{ {\ve{p}_{\xi}^{\rm A,B}}\over { m_{\xi}}}\  ,\ \ \ 
\ve{V} = {{\ve{P}}\over{M_{\rm AB}}}\  ,\ \ \ 
M_{\rm AB} = {{ M_{\rm A} M_{\rm B}}\over{M}} 
\ ,
\label{v_A,B;VAB;MAB}
\end{equation} 
\begin{equation}
\ve{l}_{\xi}^{\rm A,B} = \ve{r}_{\xi}^{\rm A,B} \times \ve{p}_{\xi}^{\rm A,B} 
\ , \ \ \ \ 
\ve{L} = \ve{R} \times \ve{P} \ .
\label{l_A,B}
\end{equation} 
The structure of the first line in Eq.~(\ref{MintClust}) is very 
similar to the structure of Eq.~(\ref{DintClust}) for 
the electric transition operator. However, the magnetic 
operator couples the cluster ``intrinsic'' EM transitions
to the relative motion between clusters in a more complicated 
way than does the corresponding electric operator. 
That is so because it also contains vector products of the subsystem 
intrinsic operators 
$\ve{d}_{\rm A,B}$ and $\ve{v}_{\rm A,B}$ with the two-cluster relative 
momentum and coordinate. 

Similarly, the magnetic spin operator of Eq.~(\ref{Mspin}) cast 
in terms of the two-cluster coordinates defined by Eq.~(\ref{rArB}) 
becomes 
\begin{equation}
\ve{M}^{\rm spin} (\ve{k}) = 
\ve{m}_{\rm A}^{\rm spin} (\ve{k})\ 
{\rm e}^{-i {{M_{\rm B}}\over{M}}\ve{k}\ve{R}} + 
\ve{m}_{\rm B}^{\rm spin} (\ve{k})\ 
{\rm e}^{i {{M_{\rm A}}\over{M}}\ve{k}\ve{R}} 
\ ,
\label{MspinClust}
\end{equation} 
\begin{equation}
\ve{m}_{\rm A,B}^{\rm spin} (\ve{k}) = 
\sum_{\xi =1}^{A,B} e {{\widehat{\mu}_{\xi}}\over {m_{\xi}}} 
\ve{s}_{\xi} {\rm e}^{-i\ve{k}\ve{r}_{\xi}^{A,B} } \ , 
\label{m_spin_A,B}
\end{equation} 
where the Coulomb gauge, given by Eq.~(\ref{CoulombGauge}), has been assumed. 

As was done in the analysis of 
the operator $\ve{D} (\ve{k})$, we assume that the magnetic 
operator defined in Eq.~(\ref{GMDM}) 
acts in a space of intrinsic states possessing a definite parity.
Also
consider the part of this operator
which gives rise to transitions that do not change parity. 
Then, in the limit $\ve{k}\rightarrow 0$,  
all but the first three terms in Eq.~(\ref{MintClust}) 
can be omitted and we can write 
\begin{equation}
\ve{M} (0)  = 
\vegr{\mu}^{\rm A} \ + \vegr{\mu}^{\rm B}\  + \ve{M}^{\rm AB} \ , 
\label{M(0)}
\end{equation} 
\begin{equation}
\vegr{\mu}^{\rm A,B}    = 
\sum_{\xi = 1}^{A,B} {{1}\over{m_{\xi}}} 
\left( {{\widehat{e}_{\xi}}\over{2}}\ \ve{l}_{\xi}^{\rm A,B} \ + 
e \widehat{\mu}_{\xi} \ve{s}_{\xi} \right)\ , 
\label{m_A,B}
\end{equation} 
\begin{equation}
\ve{M}^{\rm AB}    = 
\mu_{\rm N} C_{M1} \ve{L} \ , 
\label{MAB(0)}
\end{equation} 
\begin{equation}
C_{M1} =  {1\over{A_{\rm tot}}} 
\left( {{M_{\rm B}}\over{M_{\rm A}}} Z_{\rm A} + 
{{M_{\rm A}}\over{M_{\rm B}}} Z_{\rm B} \right) \ ,
\label{C_M1}
\end{equation} 
where $\mu_{\rm N}\equiv {{e}/{2m_{\rm N}}}$ is the nuclear magneton 
and $m_{\rm N}$ is the nucleon mass. 
Therefore, the generalised 
magnetic dipole moment determined by Eqs.~(\ref{MintClust}) 
and (\ref{MspinClust}) in the long-wavelength limit 
is the total magnetic dipole moment operator for the two-body system.  
This is a sum of the subsystem intrinsic magnetic dipole operators 
$\vegr{\mu}^{\rm A,B}$ defined in the usual way~\cite{BohrMottelson,BrussGlaud}  
with an additional contribution 
$\ve{M}^{\rm AB}$ that is due to cluster-relative orbital 
motion. Evidently, in the specific case of a two-cluster 
potential 
picture without coupling to ``intrinsic'' excitations for the subsystems, 
transition matrix elements of operators $\vegr{\mu}^{\rm A,B}$ are determined 
by the static magnetic dipole moments of the clusters A and B. 

We can proceed with the analysis of the magnetic operator
given in Eq.~(\ref{GMDM}) for low photon energies 
in a similar manner to that we have used for the electric 
operator. Disregarding  terms of order 
$O(\mu_{\rm N}k^{2})$, the part of this operator which does not change 
the cluster intrinsic parity can be written as 
\begin{equation}
\ve{M}(\ve{k}) = 
\ve{M}^{M1} - {i} \ve{M}^{M2} (\ve{k}) 
\ , 
\label{M(1)}
\end{equation} 
\begin{equation}
\ve{M}^{M1}  \equiv \ve{M} (0) 
\ , 
\label{M(0)_1}
\end{equation} 
\begin{eqnarray}
&&
\ve{M}^{M2} (\ve{k})= 
(\ve{k}\ve{R})\ \left( 
{{M_{\rm B}}\over{M}}\vegr{\mu}^{\rm A} - {{M_{\rm A}}\over{M}}\vegr{\mu}^{\rm B}  
\right) \ +\ 
{\mu_{\rm N}}  C_{M2} \{ \ve{L} , (\ve{k}\ve{R})\} 
\nonumber\\
&&
+\ {{i}\over{6}}\ve{R}\times\left( 
{{M_{\rm B}}\over{M}}\left[ H_{\rm A} , \ve{d}_{\rm A}^{E2} (\ve{k})\ \right] - 
{{M_{\rm A}}\over{M}}\left[ H_{\rm B} , \ve{d}_{\rm B}^{E2} (\ve{k})\ \right]
\right)
\nonumber\\
&&
+\ {{1}\over{3}} \left(
{{M_{\rm B}}\over{M}} \ve{d}_{\rm A}^{E2} (\ve{k}) 
- {{M_{\rm A}}\over{M}} \ve{d}_{\rm B}^{E2} (\ve{k}) 
\right)
\times \ve{V}
\ ,
\label{M2general}
\end{eqnarray} 
\begin{equation}
{d}_{{\rm A,B};\ j}^{E2}(\ve{k}) = {{1}\over{3}} \left( 
k_{l} t_{jl}^{\rm A,B} + e k_{j} Z_{\rm A,B}\ {r}_{\rm A,B}^2 \right)
\ , 
\label{d1_A,B}
\end{equation} 
\begin{equation}
C_{M2} =  {1\over{3 A_{\rm tot} M}} \left( {{M_{\rm B}^2}\over{M_{\rm A}}} Z_{\rm A} - 
{{M_{\rm A}^2}\over{M_{\rm B}}} Z_{\rm B} \right) \ ,
\label{C_M2}
\end{equation} 
where $H_{\rm A}$ ($H_{\rm B}$) and ${r}_{\rm A}^2$ (${r}_{\rm B}^2$) are 
the cluster A (B) intrinsic Hamiltonian and the operator of the
square charge radius, respectively. Details are given in Appendix~\ref{App_A}.
If we consider the case of a cluster-like potential model,
with no coupling to the intrinsic excitations and with vanishing
static electric quadrupole moments, Eq.~(\ref{M2general})  
reduces to 
\begin{eqnarray}
&&
\ve{M}^{M2} (\ve{k})= 
 (\ve{k}\ve{R})\ \left( 
{{M_{\rm B}}\over{M}}\vegr{\mu}^{\rm A} - {{M_{\rm A}}\over{M}}\vegr{\mu}^{\rm B}  
\right) \ + \ 
{\mu_{\rm N}}  C_{M2} \{ \ve{L} , (\ve{k}\ve{R})\} 
\nonumber\\
&&
+ {{1}\over{9}} e 
{{M_{\rm B} Z_{\rm A}\langle {r}_{\rm A}^2 \rangle - 
M_{\rm A} Z_{\rm B}\langle {r}_{\rm B}^2 \rangle }\over{M}} 
\ve{k} \times \ve{V}\ 
\ . 
\label{M2_frozen}
\end{eqnarray} 

Finally, note that the formalism 
outlined here paves the way for the development of an extended cluster 
model for description of low-energy photo-nuclear processes taking 
into account the internal structure and EM excitation of the clusters. 
Starting from the microscopic (single-particle) model defined by
Eqs.~(\ref{ModelRho}) and 
(\ref{ModelJ}) for the EM current operator, we have derived the low-energy EM 
transition operators, and specifically separated the terms
acting on the part of the wave function that describes the relative motion 
between clusters. 

Conversely, the cluster intrinsic structure is described by operators of 
cluster magnetic dipole and electric quadrupole moments and 
squares of charge radii. Usually, these intrinsic operators are  
defined microscopically, but the evaluation of their matrix 
elements can be done within any model, be it microscopic
(e.g., by the shell model, or by few-body techniques
for light nuclei) or macroscopic (a collective-type model). 
Therefore, our approach can be considered as a step towards construction 
of a unified framework for treatment of static electromagnetic properties 
of nuclei and low-energy nuclear radiative capture  with a link 
(as shown below) to the description of nuclear two-body scattering 
within the multichannel approach of Ref.~\cite{MCAS1}. 
Furthermore, 
this formalism can be easily generalised to the case of three-cluster 
and more complex systems through the proper extension of the set 
of the Jacobi coordinates and momenta. 
%

\section{MCAS wave functions for two-cluster states}
\label{sec4}

Consider the radiative-capture process 
\begin{equation}
{\rm A}\ +\ {\rm B}\ \rightarrow\ {\rm C}\ +\ \gamma  \ .
\label{reaction}
\end{equation} 
Initially there is a scattering state $\mid\Psi^{(+)}_{\ve{P}};
\phi_{J_{\rm A}}^{\rm A} J_{\rm A}m_{\rm A};
\phi_{J_{\rm B}}^{\rm B} J_{\rm B}m_{\rm B}\rangle$ characterised by 
the two-body c.m. momentum $\ve{P}$ and a nucleus A (B) intrinsic state 
$\phi_{J_{\rm A}}^{\rm A}$ ($\phi_{J_{\rm B}}^{\rm B}$) with the total  
spin $J_{\rm A}$ ($J_{\rm B}$) and its projection $m_{\rm A}$ ($m_{\rm B}$). 
The process forms a bound state 
$\mid \Phi_{J_{\rm C}}^{\rm C} J_{\rm C}m_{\rm C}\rangle$ which has total  
spin $J_{\rm C}$ and projection $m_{\rm C}$. 
Assume that the nuclear system is described within a two-cluster picture 
without any direct reference to single-particle coordinates and momenta. 
This conforms to the current MCAS approach~\cite{MCAS1}, which utilises 
the OPP method~\cite{Kukulin, MCAS3} to account for the Pauli principle. 
With partial-wave decomposition, the two-cluster initial- 
and final-state wave functions can be written as
\begin{eqnarray}
&&\langle\ve{R}\mid \Psi^{(+)}_{\ve{P}};
\phi_{J_{\rm A}}^{\rm A} J_{\rm A}m_{\rm A};
\phi_{J_{\rm B}}^{\rm B} J_{\rm B}m_{\rm B}\rangle = 
\sum_{{J,a,a'',}\atop{m''_{\rm A},m''_{\rm B}}}
\langle a'',m''_{\rm A},m''_{\rm B}\mid 
{\cal Y}_{J M_{J}}(\Omega_{\ve{R}},\Omega_{\ve{P}})\mid a, m_{\rm A}, m_{\rm B}\rangle 
\nonumber \\
&&\hspace*{6.0cm}
\times 
\mid \phi_{J''_{\rm A}}^{\rm A} J''_{\rm A}m''_{\rm A}\rangle 
\mid \phi_{J''_{\rm B}}^{\rm B} J''_{\rm B}m''_{\rm B}\rangle 
\ \Psi_{a'' a}^{J\ (+)}(R) 
\ ,
\label{Scattstate}
\end{eqnarray}
and
\begin{eqnarray}
\langle\ve{R}\mid \Phi^{\rm C}_{J_{\rm C}} J_{\rm C}m_{\rm C}\rangle 
&=& 
\sum_{{a',M',M'_{L},}\atop{m'_{\rm A},m'_{\rm B}}} 
C(L' M'_{L} J'_{\rm A} m'_{\rm A} J' M' J'_{\rm B} m'_{\rm B} J_{\rm C} m_{\rm C}) 
\nonumber
\\
&&\hspace*{1.0cm}
\times 
\mid \phi_{J'_{\rm A}}^{\rm A} J'_{\rm A}m'_{\rm A}\rangle 
\mid \phi_{J'_{\rm B}}^{\rm B} J'_{\rm B}m'_{\rm B}\rangle 
\ \Phi_{J_{\rm C} a'}^{\rm C}(R)\ Y_{L' M'_{L}}(\Omega_{\ve{R}}) 
\ ,
\label{Boundstate}
\end{eqnarray}
wherein we have used the short notations 
\begin{eqnarray}
&&
\mid a\rangle\equiv\ \mid P;\ \left( \left( L,J_{\rm A}\right) J''', 
J_{\rm B}\right)\ J\rangle
\ ,
\nonumber\\ 
&&\mid a'\rangle\equiv\ \mid P';\ \left( \left( L',J'_{\rm A}\right) J', 
J'_{\rm B}\right)\ J_{\rm C}\rangle\ ,
\nonumber
\\
&& 
\mid a''\rangle\equiv\ \mid P'';\ \left( \left( L'',J''_{\rm A}\right) J'', 
J''_{\rm B}\right)\ J\rangle\ ,
\label{channelnotation}
\end{eqnarray}
for the channel states. We have also made definitions 
\begin{eqnarray}
&&
\langle a'',m''_{\rm A},m''_{\rm B}\mid 
{\cal Y}_{J M_{J}}(\Omega_{\ve{R}},\Omega_{\ve{P}})\mid a, m_{\rm A}, m_{\rm B}\rangle 
\equiv 
\!\!\!\sum_{{M,M'',M''',}\atop{M_{L},M''_{L}}}\!\!\!\!\! 
C(L'' M''_{L} J''_{\rm A} m''_{\rm A} J'' M'' J''_{\rm B} m''_{\rm B} J M_{J})\  
\nonumber
\\
&&\hspace*{3.0cm}
\times 
C(L M_{L} J_{\rm A} m_{\rm A} J''' M''' J_{\rm B} m_{\rm B} J M_{J})\  
\ Y_{L'' M''_{L}}(\Omega_{\ve{R}}) \ Y_{L M_{L}}^{*}(\Omega_{\ve{P}}) \ ,
\label{somedefs1}
\\
&&
\nonumber
\\
&&
C(j_{1} m_{1} j_{2} m_{2} j_{3} m_{3} j_{4} m_{4} j_{5} m_{5} )\ 
\equiv 
\langle j_{1} m_{1} j_{2} m_{2} \mid j_{3} m_{3} \rangle 
\langle j_{3} m_{3} j_{4} m_{4} \mid j_{5} m_{5} \rangle\ . 
\label{somedefs2}
\end{eqnarray}
These formulae are given for a general case when 
non-central nuclear forces~\cite{GW} and  
inelastic ($P''\neq P$) two-body channels are to be taken into account. 
The angular-momentum coupling scheme implied 
is shown explicitly in Eqs.~(\ref{channelnotation}).

The coordinate-space scattering radial wave function $\Psi_{a'' a}^{J\ (+)}(R)$ 
has an asymptotic behaviour~\cite{Taylor} 
\begin{equation}
\Psi_{a'' a}^{J\ (+)}(R)\  _{\overrightarrow{^{R\to\infty}}} \ 
\sqrt{{2}\over{\pi}}\ {\frac{i^{L''}}{PR}}\ {{1}\over{2i}} 
\left( 
\sqrt{ {{P}\over{P''}} } S_{a'' a}^{J} O_{L''}^{(+)} (P''R) - 
\delta_{a''a} O_{L}^{(-)} (PR)\ 
\right)
\ , 
\label{Asymp1} 
\end{equation} 
where $S_{a'' a}^{J}$ is an element of the partial-wave $S$-matrix and
$O_{a}^{(-)}$ and $O_{a}^{(+)}$ are the incoming and outgoing Coulomb waves 
respectively. For the latter, we imply~\cite{LaneThomas}
\begin{eqnarray}
&&
O_{L}^{(-)} (PR)\  _{\overrightarrow{^{R\to\infty}}} \ 
\textrm{e}^{-i[ PR - \eta \ln (2PR) - \frac{L\pi}{2} ]}\ ,
\nonumber
\\
&&
O_{L}^{(+)} (PR)\  _{\overrightarrow{^{R\to\infty}}} \ 
\textrm{e}^{ i[ PR - \eta \ln (2PR) - \frac{L\pi}{2} ]}
\ , 
\label{AsympCoul1} 
\end{eqnarray} 
\begin{equation}
\eta \equiv \eta (P) = Z_{\rm A} Z_{\rm B}\alpha M_{\rm AB}/P
\ ,
\label{eta}
\end{equation} 
with $\alpha$ being the fine-structure constant. 
Next, we can write 
\begin{eqnarray}
&&
O_{L}^{(-)} (PR)\  =  
O_{L}^{R(-)} (PR)\ \ \textrm{e}^{i \sigma_{L} (P)}
\ ,
\nonumber
\\
&&
O_{L''}^{(+)} (P''R)\  = 
O_{L''}^{R(+)} (P''R)\ \ \textrm{e}^{-i\sigma_{L''} (P'')}
\ , 
\label{O_reduced} 
\end{eqnarray} 
\begin{equation}
O_{L}^{R(\pm )} (x) = 
G_{L} (x) \pm i F_{L} (x)
\ ,
\label{O_red_def}
\end{equation} 
where, as usual, for the Coulomb regular ($F_{L}$) and irregular ($G_{L}$) 
functions, one has 
\begin{eqnarray}
&&
F_{L} (PR)\  _{\overrightarrow{^{R\to\infty}}} \ 
\sin [PR - \eta \ln (2PR) - \frac{L\pi}{2}  + \sigma_{L} (P)\ ] \ ,
\nonumber
\\
&&
G_{L} (PR)\  _{\overrightarrow{^{R\to\infty}}} \ 
\cos [PR - \eta \ln (2PR) - \frac{L\pi}{2}  + \sigma_{L} (P)\ ] 
\ .
\label{AsympCoul2} 
\end{eqnarray} 
The Coulomb phase shift depends on the Sommerfeld parameter $\eta$ as 
\begin{displaymath}
\sigma_{L} (P) = \arg \Gamma (L+1+i\eta (P)\ ) \ .
\end{displaymath}

Using the definitions given in Eq.~(\ref{O_reduced}) and introducing the 
``reduced'' $S$ matrix $S_{J}^{R}$~\cite{GW} through the definition 
\begin{equation}
S_{a'' a}^{J} = 
\textrm{e}^{i\sigma_{L''} (P'')} S_{J, a'' a}^{R} \textrm{e}^{i\sigma_{L} (P)} 
\ ,
\label{S_red_def}
\end{equation} 
the asymptotic condition (\ref{Asymp1}) becomes 
\begin{equation}
\Psi_{a'' a}^{J\ (+)}(R)\  _{\overrightarrow{^{R\to\infty}}} \ 
i^{L''} \sqrt{{2}\over{\pi}}\ {\frac{\textrm{e}^{i\sigma_{L} (P)}}{\sqrt{P''P}R}}\  
{{ S_{J, a'' a}^{R} O_{L''}^{R(+)} (P''R) - 
\delta_{a''a} O_{L}^{R(-)} (PR)}\over {2i}} 
\ . 
\label{Asymp2} 
\end{equation} 

The wave function $\Psi_{a'' a}^{J\ (+)}(R)$, which is regular at $R=0$ and 
satisfies the asymptotic condition, Eq.~(\ref{Asymp2}), 
can be constructed from the MCAS multi-channel scattering matrix given in 
the Coulomb-state representation~\cite{CattPisVanz} for a separable 
two-body potential of rank $N$, namely 
\begin{equation}
V_{a''a}(p,q) = \sum_{n=1}^{N} 
\langle p \mid \chi_{a''n}\rangle \lambda_{n}^{-1}\langle \chi_{an}\mid q \rangle
\ ,
\label{SeparPot}
\end{equation} 
in which the momentum-space potential form factors are the Fourier-Coulomb 
transforms, 
\begin{equation}
\widehat{\chi}_{an} (p) \equiv \langle p \mid \chi_{an}\rangle = 
\sqrt{{2}\over{\pi}} {{1}\over{p}} \int_{0}^{\infty} 
F_{L} (pr)\ \chi_{an} (r) {\rm d}r \ ,
\label{pot_ffactors} 
\end{equation} 
of the corresponding coordinate-space form factors. 

In our present study, we employ the procedure~\cite{CantonPisent91} to construct 
the scattering radial wave function for the separable 
potential~(\ref{SeparPot}) in coordinate space.
That involves a direct resolution of 
the coupled-channel Lippmann-Schwinger equation in a finite-rank matrix 
form using the Green's function expressed through the Coulomb functions 
and for the physical outgoing solution, i.e., that obeying 
condition given in Eq.~(\ref{Asymp2}).  
Following this procedure with the MCAS scattering matrix~\cite{MCAS1}, we get 
\begin{equation}
\Psi_{a'' a}^{J\ (+)}(R) = 
i^{L''} \sqrt{{2}\over{\pi}} {{{\exp}[{i\sigma_{L}(P)}]}\over{\sqrt{P''P}R}} 
\left[
F_{L} (PR)\ \delta_{a''a}\ -\ \pi M_{\rm AB}\sqrt{P''P}\  
\Phi_{a'' a}^{J\ (+)}(R)
\right]
\ , 
\label{ScattRadwf} 
\end{equation} 
\begin{eqnarray}
&&
\Phi_{a'' a}^{J\ (+)}(R) = 
i^{L-L''} \sum_{n,n'=1}^{N} \left\{
F_{L''} (P'' R)\ \chi_{a''n}^{G} (P'',R)\ - 
G_{L''} (P'' R)\ \chi_{a''n}^{F} (P'',R)
\right. 
\nonumber 
\\
&&\hspace*{4.6cm}
\left.
+\  O_{L''}^{R(+)} (P'' R)
\ 
\widehat{\chi}_{a''n} (P'')\ 
\right\}\ 
\left[ 
\lambda - {\cal G}_{0}^{(+)}
\right]_{nn'}^{-1}
\widehat{\chi}_{a n'} (P) 
\ , 
\label{Phi(+)} 
\end{eqnarray} 
where 
\begin{eqnarray}
&&
{\chi}_{an}^{G} (P,R) \equiv 
\sqrt{{2}\over{\pi}} {{1}\over{P}} \int_{R}^{\infty} 
G_{L} (Pr)\ \chi_{an} (r) {\rm d}r
\ , 
\nonumber 
\\
&& 
{\chi}_{an}^{F} (P,R) \equiv 
\sqrt{{2}\over{\pi}} {{1}\over{P}} \int_{R}^{\infty} 
F_{L} (Pr)\ \chi_{an} (r) {\rm d}r
\ ,
\label{more_defs} 
\end{eqnarray} 
\begin{displaymath}
\widehat{\chi}_{an} (P) \equiv \chi_{an}^{F} (P,0)\ .
\end{displaymath}
The quantity in the square brackets of 
Eq.~(\ref{Phi(+)}) is a matrix, with its elements in the space of 
separable-decomposition indices being 
\begin{eqnarray}
&&
\left[ {\cal G}_{0}^{(+)} \right]_{n n'} = 2 M_{\rm AB}\ 
\left[ \ 
\sum_{a=1}^{\rm open} \int_{0}^{\infty} 
\widehat{\chi}_{a n} (x) 
{{x^{2}}\over{P^{2} -x^{2} + i0}} 
\widehat{\chi}_{a n'} (x) \ {\rm d}x\ 
\right.
\nonumber 
\\
&&\hspace*{3.0cm}
\left. 
-
\sum_{a}^{\rm closed} \int_{0}^{\infty} 
\widehat{\chi}_{a n} (x) 
{{x^{2}}\over{h_{a}^{2} +x^{2} }} 
\widehat{\chi}_{a n'} (x) \ {\rm d}x\ 
\right] 
\ , 
\label{etagreenG0}
\end{eqnarray}
where
\begin{equation}
\left[ \lambda \right]_{nn'} = \lambda_{n} \delta_{nn'} 
\ , 
\label{etagreenlambda} 
\end{equation} 
\begin{equation}
P = \sqrt{2M_{\rm AB} (E_{\rm c.m.} - E_{a}^{\rm th})}\ , \ \ \ 
h_{a} = \sqrt{2M_{\rm AB} (E_{a}^{\rm th} - E_{\rm c.m.})} 
\ , 
\label{channelenergy} 
\end{equation} 
where $E_{a}^{\rm th}$ is the threshold energy of channel $a$
and $E_{\rm c.m.}$ is the elastic-channel kinetic energy. Both are 
considered in the c.m. frame. 
The function given in Eq.~(\ref{Phi(+)}) oscillates at large radii
with amplitude given by the multi-channel $T$ matrix. Thus we can write 
\begin{equation}
O_{a'' a}^{R(+)}(\infty )\ T_{a'' a}^{J} = \Phi_{a'' a}^{J\ (+)}( \infty ) 
\ , 
\label{Phi_infty} 
\end{equation} 
\begin{equation}
S_{J, a'' a}^{R} = 
\delta_{a'' a}\ -\ 2 \pi i M_{\rm AB} \sqrt{P'' P}\ T_{a'' a}^{J} 
\ , 
\label{S_matrix} 
\end{equation} 
where the  asymptotic outgoing Coulomb wave $O_{a'' a}^{R(+)}(\infty )$ is given by 
Eqs.~(\ref{AsympCoul1}) and (\ref{O_reduced}).

Thus, we have constructed the radial wave function for the two-fragment 
initial (scattering) state, which is general in the sense that 
possibilities of coupling to inelastic channels are 
taken into consideration along with coupling between elastic channels 
due to the non-central part of the two-body potential. 
The wave function construction for bound states can be fulfilled 
through solving the coupled-channel Schr{\"o}dinger problem 
analogously to the procedure outlined above for the scattering wave function. 
That is so as the MCAS approach utilises sturmian expansions to construct 
the separable interactions, with which
 determination of both the positive-energy (resonance) and 
negative-energy (bound) two-body states solutions of the Lippmann-Schwinger
equations can be found based upon analytic properties of 
the sturmian functions. As all details have been given~\cite{MCAS1}, 
we do not reproduce them here. 
We conclude this section by noting that the inversion of the matrix 
$\left[ \lambda - {\cal G}_{0}^{(+)} \right]$ contains all the dynamical 
features of 
the two-cluster interaction process, including possible formation of a compound 
system at the given resonant energy. 


\section{Reaction ${^{3}{\rm He}}({\alpha},\gamma ){^{7}{\rm Be}}$}
\label{sec5}

The formalism outlined above encompasses studies of EM transitions 
in which the role of the finite sizes and collective excitations 
of the fragments can be taken into account. 
However, for an initial application of the formalism, we have chosen 
the reaction ${^{3}{\rm He}}({\alpha},\gamma ){^{7}{\rm Be}}$.
In this case there are no low-energy
inelastic two-body channels, whereas the elastic channels are uncoupled. 
Such a choice 
has been governed in part by the amount of experimental information available 
as well as the many 
theoretical studies made of it in the past. Furthermore, 
a goal of this illustrative calculation was to establish the degree 
of applicability of the MCAS model in a description 
of EM processes. 
We also wish to compare theoretical ingredients of this calculation 
with those given by other approaches. In the first instance we have not
sought to achieve a quantitative agreement 
with experimental data. Thus we have not varied parameters of the two-body 
potential in our calculation. Instead, the radial wave function's for 
the initial and final states were constructed using the nuclear potential of 
Ref.~\cite{MCAS2} which had been found to best reproduce 
the states and resonance widths of $^7$Li. 

The differential cross section for the radiative-capture 
process can be written as 
\begin{eqnarray}
&&
{{{\rm d}\sigma^{\rm c.m.}}\over{{\rm d}\Omega _{\gamma}}} = 
4 {\pi}^{2}\alpha\ {{M_{\rm AB}}\over{\mid\ve{P}\mid}} 
{{M }\over{M +E_{\gamma}}}\ E_{\gamma}^{3}\ 
\left( 
\mid \widetilde{\ve{D}}_{if}\mid^{2} - 
\mid \widehat{\ve{k}} \widetilde{\ve{D}}_{if}\mid^{2} +
\mid \widetilde{\ve{M}}_{if} \times \widehat{\ve{k}} \mid^{2}  
\right.
\nonumber
\\ 
&&\hspace*{7.0cm}
\left.
+ 
2 \textrm{Re} \left( \widetilde{\ve{D}}_{if}^{*} 
\left[ \widetilde{\ve{M}}_{if} \times \widehat{\ve{k}} \right]\right)\  
\right)
\ , 
\label{sigma_diff}
\end{eqnarray} 
\begin{displaymath}
e\ \widetilde{\ve{D}}_{if}\equiv \langle f\mid \ve{D}(\ve{k})\mid i \rangle , \ \ 
e\ \widetilde{\ve{M}}_{if}\equiv \langle f\mid \ve{M}(\ve{k})\mid i \rangle , \ \ 
\widehat{\ve{k}} \equiv {{\ve{k}}\over{\mid\ve{k}\mid}}\ .
\end{displaymath}
All the quantities are in  the c.m. frame and 
averaging (summation) over initial (final) nuclear system spin states 
is implied. 
Details for calculation of the matrix elements 
$\langle f\vert \ve{D}(\ve{k})\vert i \rangle$ 
and 
$\langle f\vert \ve{M}(\ve{k})\vert i \rangle$ 
are given in Appendix~\ref{App_B}. 
When calculating the magnetic transition matrix element,  
only the first two terms of the $M2$ operator, Eq.~(\ref{M2general}), 
are retained. 
The last term of that formula is negligibly small for the energy 
range covered by the calculation while  the third one is zero for the 
reaction considered. 

To calculate the radial overlap integrals for 
the ${^{3}{\rm He}}({\alpha},\gamma ){^{7}{\rm Be}}$ reaction,  
scattering- and bound-state radial wave functions have been found 
following the MCAS formalism 
described in Sec.~\ref{sec4}. To minimize any impact of 
computational uncertainties in evaluation of the bound-state wave functions 
upon the results, we replaced these wave functions at a very long two-cluster 
distance ($\mid\ve{R}\mid >10$~fm) by the properly renormalised Whittaker 
functions 
that correspond to experimentally known binding energies. 
The radial wave function 
for the 
{$^7$Be} ground state gives rise to the rms charge radius of 2.61~fm, 
which is in 
fair agreement with the experimental value of 
$2.52 \pm 0.3$~fm~\cite{Tanihata}. 
The Coulomb wave functions 
were calculated using the {\sc cernlib} program~\cite{ThompsonBarnett}.

\begin{figure}[hbtp]
    \resizebox{11cm}{!}{\includegraphics{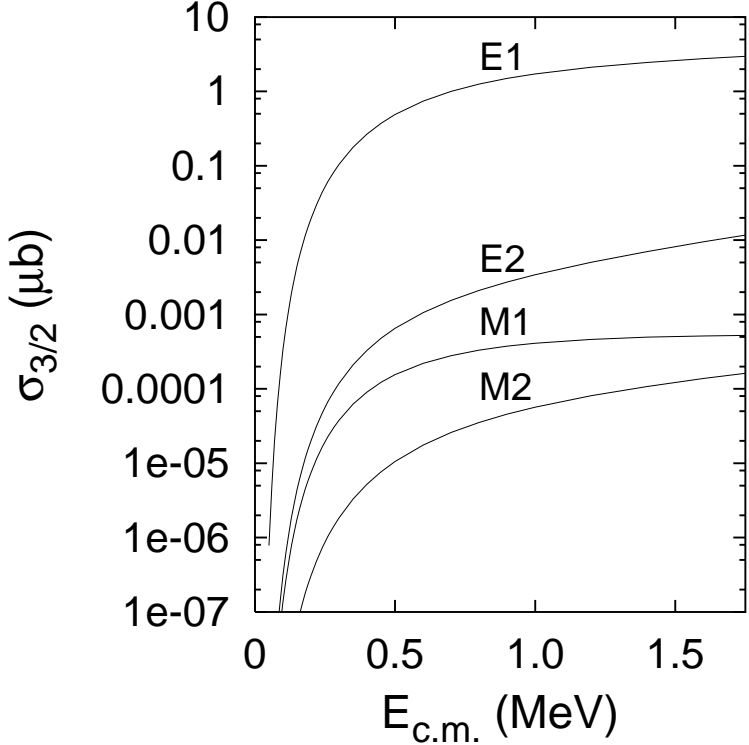}}
\caption{Multipole contributions to the total cross section 
for ${}^3$He-${}^4$He capture leading to the ground state of  
$^{7}$Be.
}
    \label{fig1}
\end{figure}
The main contribution to the ${^{3}{\rm He}}({\alpha},\gamma ){^{7}{\rm Be}}$ 
reaction amplitude at low energies is well known to result from the electric dipole 
transition. This is so because the quantum numbers of both the {$^7$Be} ground 
state (${{3}\over{2}}^{-}$) and its excited (${{1}\over{2}}^{-}$, 429~keV) state allow 
the electric dipole transition to occur from the scattering $s$-wave. 
On the other hand, the operator in Eq.~(\ref{D_E1}) responsible for 
this transition in this case is neither vanishing nor small. 
As a consequence, the angular distribution obtained 
is rather flat at very low two-cluster relative energies. Some small 
non-uniformity 
is caused by the interference between the dipole transitions 
from initial $s$ and $d$ waves. At higher energies, the $E2$ transition 
due to the operator given in Eq.~(\ref{TAB}) also plays a role, 
making the c.m. cross section 
asymmetric about {90$^{\circ}$}. 
Magnetic transitions arising from the operator in Eq.~(\ref{M(1)}) 
are only
minor contributions to the cross section as shown in Fig.~\ref{fig1}. 
These observations agree with the conclusions given in Ref.~\cite{Kim81}. 

\begin{figure}[hbtp]
    \resizebox{15cm}{!}{\includegraphics{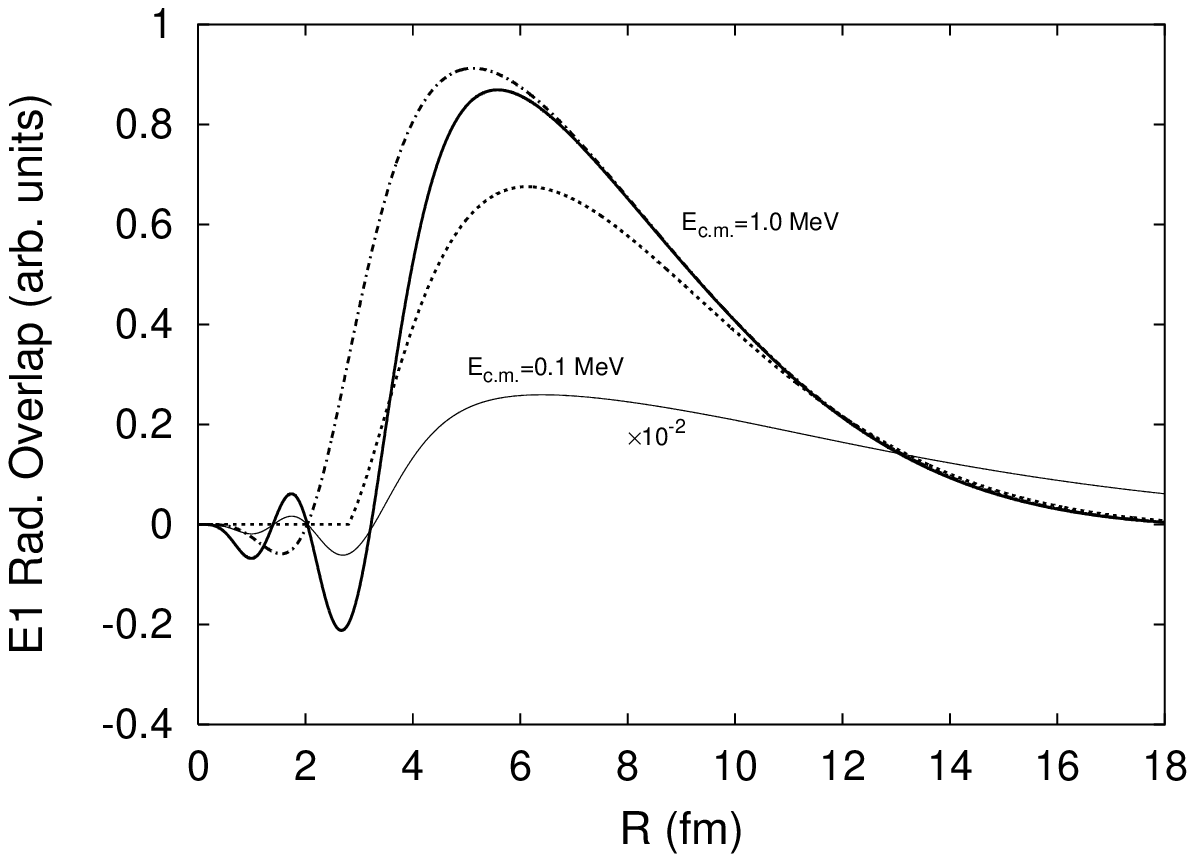}}
\caption{The real part of the radial overlap function for the electric dipole 
transition to the {$^{7}$Be} ground state from an initial relative $s$-wave. 
The thick and thin solid curves were obtained with the scattering radial 
wave function defined in Eq.~(\protect\ref{ScattRadwf}) for c.m. energies of 
1~MeV and 100~keV, respectively. 
The dotted and dot-dashed lines are the overlaps found using the 
repulsive hard-core (pure Coulomb) initial-state wave functions for a 
c.m. energy of 1~MeV.}
    \label{fig2}
\end{figure}
The real parts of the radial overlap functions (Eq.~(\ref{overlap_def}) 
in  Appendix~\ref{App_B}) for the dominant electric dipole transition
for the captures to the {$^{7}$Be} ground state from the
relative $s$-state at   
c.m. energies of 100~keV and 1~MeV are shown 
in Fig.~\ref{fig2}. For convenience, the Coulomb-phase exponential factor 
in Eq.~(\ref{ScattRadwf}) has been omitted. 
The initial- and final-state radial wave functions constructed within the 
MCAS scheme  exhibit
oscillations in the radial overlap at small ($<4$~fm) distances. 
These wave functions and the resulting overlap function are in remarkable 
agreement 
with those given by the OCM and potential-model calculations~\cite{Kim81}. 
In Fig.~\ref{fig2} we also compare those results with the overlaps 
when the initial-state radial wave function 
is a pure Coulomb function and the scattering wave function is that for 
the repulsive hard-core (RHC) model~\cite{Tombrello}. The hard sphere 
radius was taken as 2.8~fm. These overlap functions indicate that the 
photon emission 
occurs, on average, at very large extra-nuclear distances. 
However, as seen also in Fig.~\ref{fig2}, taking into 
account the nuclear forces leads to a significant repulsive 
effect which is very important for quantitative estimates. 
Comparison of the curves for $E_{\rm c.m.}=$1~MeV and 100~keV shows 
the degree of spread of the EM interaction area with the decrease in 
the two-cluster relative energy.  

\begin{figure}[hbtp]
    \resizebox{15cm}{!}{\includegraphics{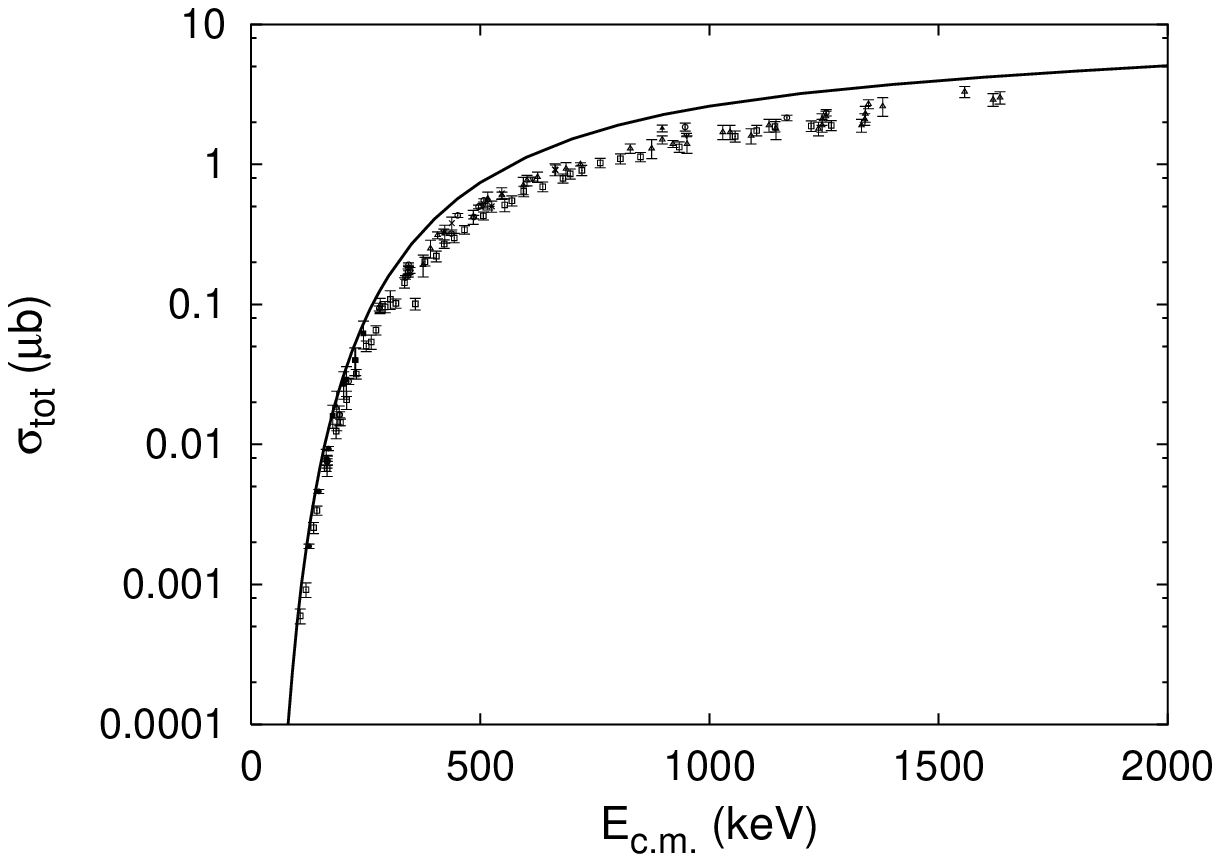}}
\caption{Total cross section of ${^{3}{\rm He}}({\alpha},\gamma ){^{7}{\rm Be}}$ 
reaction. The filled circles at $E_{\rm c.m.}=$ 127, 148 and 169~keV are 
recent LUNA data~\protect\cite{Bemmerer}. Other experimental points were 
taken from 
Refs.~\protect\cite{ParkerKavanagh,Alexander,Hilgemeier,Krawinkel,
Nagatani,Osborne,Robertson,Singh}. 
}
    \label{fig3}
\end{figure}

\begin{figure}[hbtp]
    \resizebox{15cm}{!}{\includegraphics{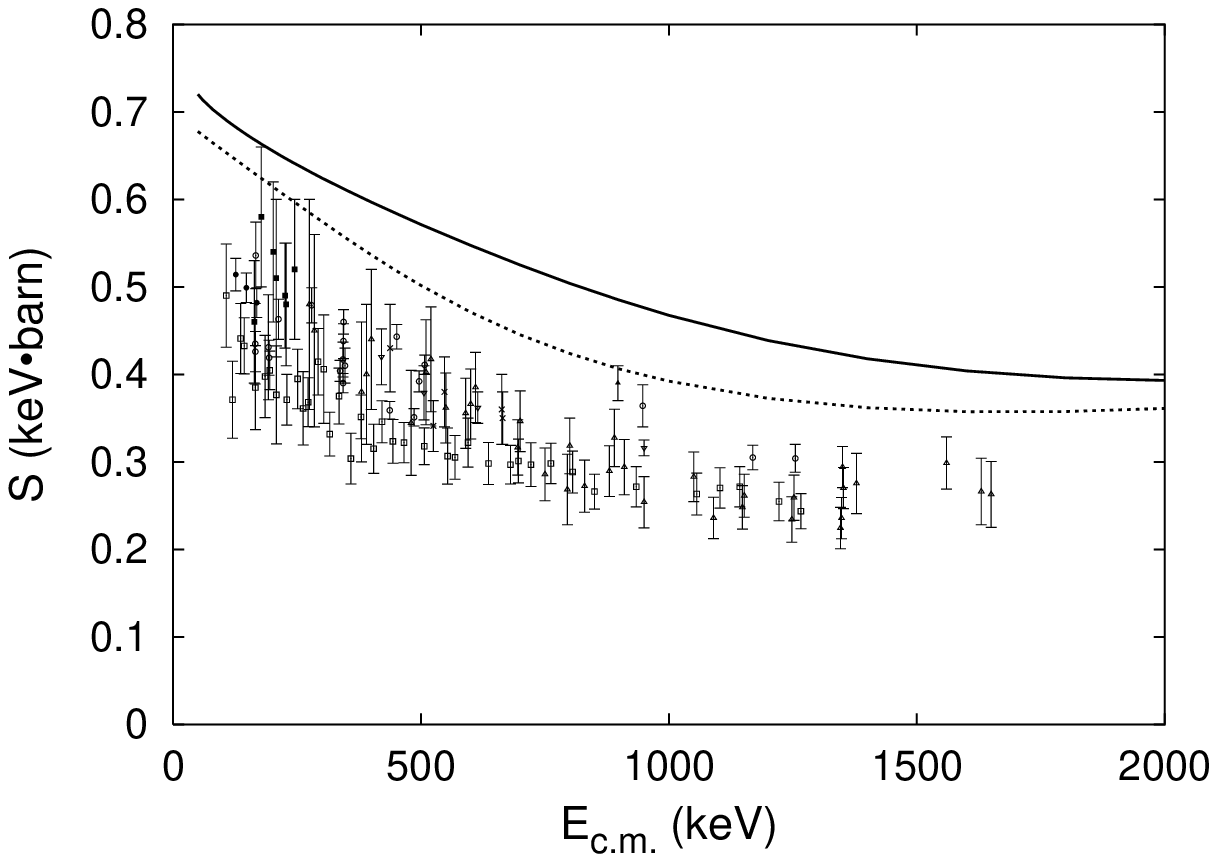}}
\caption{Astrophysical $S$ factor for reaction 
${^{3}{\rm He}}({\alpha},\gamma ){^{7}{\rm Be}}$. 
The notation is as used in Fig.~\protect\ref{fig3}. The dotted curve 
displays the result obtained with the initial-state radial wave function 
constructed for the RHC model.
}
    \label{fig4}
\end{figure}

In Figs.~\ref{fig3} and \ref{fig4} respectively we show 
our results for the total cross section 
$\sigma_{\rm tot}$ and for the corresponding astrophysical $S$-factor 
which is defined as  
\begin{equation}
S (E_{\rm c.m.}) = 
E_{\rm c.m.} \sigma_{\rm tot} (E_{\rm c.m.}) 
\exp{\left[ 2\pi\eta (P)\right]} 
\ .
\label{S-factor}
\end{equation}
Here,
\begin{displaymath}
P = \sqrt{2 M_{\rm AB} E_{\rm c.m.} }
\ .
\end{displaymath}
Our calculated results reproduce the energy dependence for the 
${^{3}{\rm He}}({\alpha},\gamma ){^{7}{\rm Be}}$ cross section and 
for the $S$-factor very well. However, the calculated cross sections are
$\sim$40\%  larger than the data. This 
is typical~\cite{Kim81,WLKT83} of both the potential 
and microscopic RGM cluster-model calculated results. 
How this 
overestimate depends upon the choice of the nuclear potential has
been studied by Kajino~\cite{Kajino}. 
He discovered that the RGM calculation with a particular nuclear force,
{\em viz.}, 
the MHN interaction~\cite{Tanabe}, could reproduce the absolute data 
for the ${^{3}{\rm He}}({\alpha},\gamma ){^{7}{\rm Be}}$ cross section and 
astrophysical $S$ factor that were known at that time. 
Nevertheless, the $S$-factor extrapolation to the zero energy 
$S(0)=0.50\pm 0.03$~keV$\cdot$barn 
suggested in Ref.~\cite{Kajino}  
is smaller than $S(0)=0.547\pm 0.017$~keV$\cdot$barn given by 
the latest $R$-matrix fit to all current data, including 
that from the LUNA experiment~\cite{Bemmerer}. 
Evidently, the two-cluster $\alpha + ^{3}$He configuration 
does not exhaust all the possibilities for the $^7$Be bound state~\cite{MCAS2}.
In principle, that could explain why the magnitude of the calculated 
cross section is larger than experiment. 
However, it was indicated~\cite{Langanke00} that extension of 
the two-cluster model space for $^7$Be by inclusion of the p$+^{6}$Li 
channel might even make the agreement between the theory and experiment
worse. 

In contrast, the variational Monte Carlo calculation~\cite{Nollett01} 
made with the two-body wave functions constructed from 
nucleon-nucleon interactions found an $S$ factor that lay
significantly below the experimental values 
for the ${^{3}{\rm He}}({\alpha},\gamma ){^{7}{\rm Be}}$ reaction. 
In particular, the value of 
$S(0)=0.40$~keV$\cdot$barn has been extrapolated. 
But this approach, without the inclusion of multi-nucleon forces,
was not able to reproduce satisfactorily the $^7$Be 
bound-state levels and the binding energies for the  separated
$^3$He and $^4$He fragments. 
In addition, phenomenological procedures had to be used to 
construct the two-cluster (7-body) scattering wave functions. 
With respect to this last point,  note that recent 
few-body techniques have been developed to deal with this difficult 
problem~\cite{LIT}.  

Though our treatment is not of the ``ab initio'' type, 
both the initial- and final-state radial wave functions 
have been constructed using the same two-body potential~\cite{MCAS2};
that potential 
with which 
the $^{7}$Li two-cluster binding energy and a number of resonance positions 
and widths were determined. This potential also gave 
the $\alpha + ^{3}$He system resonance parameters
listed in Table~1 and they, too, are in good 
agreement with experiment. In this table, the
state positions are given relative to $\alpha-{}^{3}$He breakup threshold
and all units are MeV.
\begin{table}[hbp]
\begin{center}
\caption{Spectrum of states for ${}^{7}$Be as $\alpha + {}^{3}$He system.}
\begin{tabular}{ccc}
\hline
\hline
$J^{\pi}$ & Level position\ (width) & Exp. level position\ (width)\\
\hline
${{3}\over{2}}^{-}$ & -1.53         & -1.59 \\
${{1}\over{2}}^{-}$ & -0.84         & -1.16 \\
${{7}\over{2}}^{-}$ &  3.07\ (0.180) & 2.98\ (0.175) \\
${{5}\over{2}}^{-}$ &  5.09\ (1.2)   & 5.14\ (1.2) \\
\hline
\hline
\end{tabular}
\end{center}
\label{I}
\end{table}
Nevertheless, the $^{7}$Be ${{1}\over{2}}^{-}$ excited state appears to be 
too weakly bound in our calculations. This affects the corresponding 
bound-state wave function behaviour in the EM interaction region. As a result, 
we obtained the branching ratio (ratio of cross section for capture into 
the ${{1}\over{2}}^{-}$ state to that for transition into the ground state) 
equal to 0.5. That is slightly above the average experimental value of 
$\sim$0.45.  Our calculated ratio essentially
is independent of the energy in 
complete agreement with both experimental data and other calculations. 

Using the MCAS two-cluster potential~\cite{MCAS2} 
the existing data on ${^{3}{\rm H}}({\alpha},\alpha ){^{3}{\rm H}}$ and 
${^{4}{\rm He}}({^{3}{\rm He}},{^{3}{\rm He}} ){^{4}{\rm He}}$ scattering 
reactions for the c.m. energy range between 2 and 7~MeV 
is well reproduced. It also gives good representation of the 
low lying states in the mass-7 compound nuclei that can be formed. 
When choosing the ``optimal'' potential parameters in that treatment, 
there were no evaluation of the nuclear scattering phase shifts at lower 
energies. 
Comparison of our result for the astrophysical $S$ factor with that 
obtained in case of replacement of the MCAS initial-state radial wave function 
by the RHC scattering wave function (see Fig.~\ref{fig4}) suggests
 that the agreement of 
our calculations with experimental data could be somewhat improved 
by correction of 
the MCAS potential parameters for better description of 
nuclear scattering at very low (below the lowest resonance) energies. 
Though, one should keep in mind, when performing this comparison, that 
the RHC initial-state radial wave function is not properly normalized. 

Adjustment of the potential given in Ref.~\cite{MCAS2} to find 
a better fit to the ${^{3}{\rm He}}({\alpha},\gamma ){^{7}{\rm Be}}$ radiative 
capture data has not been attempted in this study as our primary
goal has been to present a comprehensive formalism for the treatment 
of radiative-capture reactions within the MCAS two-cluster picture.
We have made a first application of such an approach, but only in a simplified 
test case. 
The general approach has been developed, however, with a view of
application to  
radiative-capture processes involving more complex nuclei.
For this reason we paid attention to particular aspects 
of the EM transition such as the role of finite fragment sizes and 
collective (rotation and vibration) excitations. 
Nevertheless, with the MCAS scheme and without any 
extra adjustment of the given potential model we could reproduce 
the ${^{3}{\rm He}}({\alpha},\gamma ){^{7}{\rm Be}}$ cross section 
as well as other theoretical models, be they 
microscopic or based on a potential specifically
constructed to describe the capture cross section.

\section{Conclusions and outlook}
\label{sec6}

A formalism has been presented to describe direct radiative-capture 
reactions at low energies within an extended two-cluster potential 
model. Unlike customary potential cluster models 
in which the interacting nuclear fragments are usually treated 
as point-like objects, the EM current operator has been used so that 
finite sizes 
of clusters and their intrinsic excitations are taken into account. 
Inherent in the approach is the explicit gauge independence and 
reproduction of the low-energy behaviour of the EM transition 
amplitude, which 
follows from the current conservation {\em viz.}, fulfilment 
of the Siegert theorem~\cite{Siegert} for electric transitions in 
the long-wavelength limit. These properties are ensured by 
extension~\cite{LevSh93,FriarFall} of the Siegert theorem which 
served as a starting point to build the photon-emission operator. 
The formalism was used to construct the electric and magnetic operators 
responsible for low-energy EM transitions in which the intrinsic parities 
of clusters remain unchanged. 
Starting with a microscopic single-nucleon current model, 
construction of these operators lead 
to expressions in terms of 
macroscopic quantities which refer to fragment EM static properties or 
intrinsic transitions such as the cluster mean-square radius and 
magnetic dipole and electric quadrupole moments. 
Such a feature allows one to perform calculations with wave functions 
of a nuclear model including collective-type coupled-channel dynamics. 

Indeed, it has been shown how 
the photon-emission operator can be 
utilised to calculate the cross section of the low-energy radiative-capture 
process~(\ref{reaction}) using multi-channel radial wave functions for
the initial 
(scattering) and final (bound) states. One of the goals of this 
study was to match the constructed EM current with the MCAS coupled-channel 
approach~\cite{MCAS1,MCAS2} developed for description of nuclear scattering. 
We have presented the explicit construction for the scattering radial
wave function 
based on the MCAS formalism. The corresponding expression is given for 
a separable two-body potential of an arbitrary finite rank in terms 
of the potential form factors, the MCAS $T$ matrix and the regular and 
irregular Coulomb functions. This wave function construction is general 
in the sense 
that it is given for a two-cluster nuclear potential with non-central 
forces. Possible inelastic channels are taken into account as is
Coulomb distortion. Detailed formulae 
for the matrix elements of the EM operator between the constructed 
nuclear states have also been given. 
Thus, applicability of the MCAS 
theory has been extended to cover nuclear radiative-capture reactions 
including ones of astrophysical relevance. 

Finally, we have applied the formalism to calculate 
the  total cross section and astrophysical $S$ factor for 
the ${^{3}{\rm He}}({\alpha},\gamma ){^{7}{\rm Be}}$ radiative capture at low 
energies. This reaction is strongly dominated by the electric dipole 
transition, with the effects due to cluster sizes and intrinsic structure 
being negligible. In addition, there are no inelastic channels in 
the $\alpha + {^{3}{\rm He}}$ system, while the elastic channels are 
uncoupled. Thus many features of the developed formalism are not
required in analysing this reaction. However, not only does this
application serve our purpose, the
reaction itself is important from the astrophysical 
point of view.  For the latter reason it has been extensively 
studied both theoretically (in different models) and experimentally. 
Our calculation 
aimed at establishing the level at which the MCAS two-body potential scheme 
is able to reproduce the observed cross-section (and $S$-factor) energy 
dependence at low energy without adjustment of potential 
parameters. To obtain the ${^{3}{\rm He}}({\alpha},\gamma ){^{7}{\rm Be}}$ 
cross section, we used the MCAS potential~\cite{MCAS2} with parameters 
providing the best fit to state positions and resonance widths for 
the mirror two-fragment system $\alpha + {}^{3}$H. 
The energy dependence of our calculated cross section 
is in a fair agreement with experiment 
and with that found using other theoretical models. 
The calculated $S$-factor 
is about 40~\% above the experimental data, a result which is also
typical of calculations performed within other models,
in particular those based upon the RGM. 

As shown in Ref.~\cite{Kajino},
better agreement with experiment can be achieved 
if one employs a two-body 
potential with parameters tuned to describe fairly well not only 
the two-fragment bound-state levels and resonance parameters, but also 
the static EM properties of the final nucleus. However, 
the corresponding potential analysis should also include 
the correction to the nucleus static magnetic moment due to 
the interaction current contribution. That comes from nonlocalities 
(dependence on the velocity) of the two-cluster potential. 
Besides the ``one-body'' 
EM operator discussed in the present paper,  
a procedure for constructing the interaction current for the nuclear 
two-fragment system is needed. That is work in progress.

The formalism presented herein,
based on the construction of the EM operator
and the wave functions built using the MCAS approach~\cite{MCAS1,MCAS2}, 
establishes a single potential framework for a unified 
description of low-energy nuclear scattering, radiative-capture
reactions, and of 
static EM properties of nuclei, in the case that they can be treated 
as two-cluster systems.  It is our intention to
consider scattering and capture reactions
involving more complex nuclei than the relatively simple ${}^{3,4}$He. 
In those future studies, we
expect that the role of cluster 
sizes and intrinsic excitations in forming the energy dependence for 
the corresponding astrophysical $S$ factors will be of great importance. 

The authors express their gratitude to K. Amos, for critical reading of 
the manuscript. 
L.G.L. would like to thank the University of Padova and I.N.F.N. 
(section Padova) for the kind hospitality and support in 2006~-- 2007.  

\appendix 
\section{Appendix: Details of the magnetic operator}
\label{App_A}

Assume that nuclear system dynamics is described by a Hamiltonian 
\begin{equation}
H = H_{\rm A} + H_{\rm B} + H_{\rm AB}\ , 
\label{Hamiltonian}
\end{equation} 
with $H_{\rm A}$ ($H_{\rm B}$) being the subsystem A (B) intrinsic 
Hamiltonian, and $H_{\rm AB}$ describing the two-cluster relative 
motion. 
Let us obtain the $k$-order correction to the leading-order 
contribution~(\ref{M(0)}) in the part of the magnetic operator~(\ref{GMDM}), 
which gives rise to transitions without changing the parity of the 
cluster intrinsic states. 

First, consider the joint contribution of order ${\mu}_{\rm N}{k}$ 
given by the first and sixth terms of Eq.~(\ref{MintClust}), 
for which we have
\begin{equation}
\ve{X}_{16} (\lambda\ve{k}) = 
-i \lambda {{M_{\rm B}}\over{M}}\left(
(\ve{k}\ve{R})\ \ve{m}_{\rm A}^{\rm orb} (0) \ + \ 
\ve{R} \times {{1}\over{2}}\sum_{\alpha =1}^{A} 
\widehat{e}_{\alpha} \{ \ve{v}_{\alpha}^{\rm A}, \ve{k} 
\ve{r}_{\alpha}^{\rm A}\}
\ \right)\ . 
\label{X_16}
\end{equation} 

If the intrinsic potential for the cluster A is local, then the relationship 
\begin{equation}
\ve{v}_{\alpha}^{\rm A} =  
i \left[ H_{\rm A} , \ve{r}_{\alpha}^{\rm A} \right]
\label{v_A=i[H,r_a]}  
\end{equation} 
is equivalent to the one given for the quantities $\ve{v}_{\alpha}^{\rm A}$ 
and $\ve{p}_{\alpha}^{\rm A}$ by Eq.~(\ref{v_A,B;VAB;MAB}). 
However, Eq.~(\ref{v_A=i[H,r_a]}) 
can be considered as a definition of operator $\ve{v}_{\alpha}^{\rm A}$ 
which enters Eq.~(\ref{MintClust}) in a more general case 
(see Ref.~\cite{BohrMottelson}, p.~393).

Using Eq.~(\ref{v_A=i[H,r_a]}), we get the relationship 
\begin{eqnarray}
&&
i \left[ H_{\rm A} , \ve{d}_{\rm A} (\lambda\ve{k} )\right] = 
i \sum_{\alpha =1}^{A} \widehat{e}_{\alpha} \left[  H_{\rm A} , 
\ve{r}_{\alpha}^{\rm A} {\rm e}^{-i\lambda\ve{k}\ve{r}_{\alpha}^{\rm A}}\right] 
\nonumber\\
&&
= 
\sum_{\alpha =1}^{A} \widehat{e}_{\alpha}
\ve{v}_{\alpha}^{\rm A} {\rm e}^{-i\lambda\ve{k}\ve{r}_{\alpha}^{\rm A}}\ 
+ i \sum_{\alpha =1}^{A} 
\widehat{e}_{\alpha} \ve{r}_{\alpha}^{\rm A} \left[ H_{\rm A} ,
{\rm e}^{-i\lambda\ve{k}\ve{r}_{\alpha}^{\rm A}}\right]  
\nonumber\\
&&
= \ve{v}_{\rm A}(0)\ - i \lambda \sum_{\alpha =1}^{A} \widehat{e}_{\alpha} 
\ve{v}_{\alpha}^{\rm A} \ (\ve{k}\ve{r}_{\alpha}^{\rm A}) - 
i \lambda \sum_{\alpha =1}^{A} \widehat{e}_{\alpha} \ve{r}_{\alpha}^{\rm A}\ 
(\ve{k}\ve{v}_{\alpha}^{\rm A})\ + O({\mu}_{\rm N}{k}^{2})\ .
\label{Comm.1}  
\end{eqnarray} 
On the other hand, we can write for the quantities $\ve{d}_{\rm A,B}$ 
\begin{equation}
\ve{d}_{\rm A,B} (\lambda\ve{k}) = 
\ve{d}_{\rm A,B} (0) - i\lambda \ve{d}_{\rm A,B}^{(1)} (\ve{k}) 
+ O({k}^{2}) \ ,
\label{d_{A,B}_expansion}  
\end{equation} 
\begin{equation}
{d}_{{\rm A,B};\ j}^{(1)}(\ve{k}) = {{1}\over{3}} \sum_l \left( 
k_{l} t_{jl}^{\rm A,B} + e k_{j} Z_{\rm A,B} {r}_{\rm A,B}^2 \right)
\ , 
\label{d1_A,B_appendix}
\end{equation} 
\begin{displaymath}
(j,l = 1,2,3)
\end{displaymath}
where $t_{jl}^{\rm A,B}$ is the cluster A, B electric quadrupole 
operator defined by Eq.~(\ref{tA,B}), and
\begin{equation}
{r}_{\rm A,B}^2 =  {{1}\over{Z_{\rm A,B}}} 
\sum_{\xi =1}^{A,B} 
\widehat{e}_{\xi} \left[ {\ve{r}_{\xi}^{\rm A,B}}\right]^{2} 
\label{r2_A,B}
\end{equation} 
is the square charge radius operator for the corresponding subsystem. 
Due to the parity selection, the contributions from $\ve{v}_{\rm A}(0)$ 
and $\ve{d}_{\rm A}(0)$ vanish for transitions between same-parity 
$H_{\rm A}$ eigenstates. 
Hence, owing to Eqs.~(\ref{Comm.1}) and (\ref{d_{A,B}_expansion}), 
we can write 
\begin{equation}
\sum_{\alpha =1}^{A} \widehat{e}_{\alpha} 
\ve{v}_{\alpha}^{\rm A} \ (\ve{k}\ve{r}_{\alpha}^{\rm A}) = 
- \sum_{\alpha =1}^{A} \widehat{e}_{\alpha} 
\ve{r}_{\alpha}^{\rm A} \ (\ve{k}\ve{v}_{\alpha}^{\rm A}) 
+ i\left[ H_{\rm A} , \ve{d}_{\rm A}^{(1)} (\ve{k})\ \right] 
\ . 
\label{Rel.1}  
\end{equation} 
omitting higher-order contributions. 
The latter relationship allows us to write 
\begin{eqnarray}
&&
{{1}\over{2}}\sum_{\alpha =1}^{A} 
\widehat{e}_{\alpha} \{ \ve{v}_{\alpha}^{\rm A}, 
\ve{k} \ve{r}_{\alpha}^{\rm A}\} = 
{{1}\over{2}}\sum_{\alpha =1}^{A} 
\widehat{e}_{\alpha} (\ve{k} \ve{r}_{\alpha}^{\rm A})\ \ve{v}_{\alpha}^{\rm A}
-
{{1}\over{2}}\sum_{\alpha =1}^{A} 
\widehat{e}_{\alpha} \ve{r}_{\alpha}^{\rm A}\ (\ve{k} \ve{v}_{\alpha}^{\rm A})\ 
+ {{i}\over{2}}\left[ H_{\rm A} , \ve{d}_{\rm A}^{(1)} (\ve{k})\ \right] 
\nonumber\\
&&
= -{{1}\over{2}} \sum_{\alpha =1}^{A} \widehat{e}_{\alpha} 
\ve{k}\times \left[ \ve{r}_{\alpha}^{\rm A}\times \ve{v}_{\alpha}^{\rm A}\right]
+ {{i}\over{2}}\left[ H_{\rm A} , \ve{d}_{\rm A}^{(1)} (\ve{k})\ \right] 
\ .
\label{Rel.2}  
\end{eqnarray} 
Substituting Eq.~(\ref{Rel.2}) into Eq.~(\ref{X_16}) and omitting longitudinal 
terms i.e. those proportional to vector $\ve{k}$, we get 
\begin{equation}
\ve{X}_{16} (\lambda\ve{k}) = 
-i {{3}\over{2}}\lambda {{M_{\rm B}}\over{M}} 
(\ve{k}\ve{R})\ \ve{m}_{\rm A}^{\rm orb} (0) \ + \ 
{{1}\over{2}}\lambda {{M_{\rm B}}\over{M}} \ve{R} \times 
\left[  H_{\rm A} , \ve{d}_{\rm A}^{(1)} (\ve{k})\ \right] 
\ . 
\label{X_16_result}
\end{equation} 
Analogously, the joint contribution from the second and seventh terms 
of Eq.~(\ref{MintClust}) is given by
\begin{equation}
\ve{X}_{27} (\lambda\ve{k}) = 
i {{3}\over{2}}\lambda {{M_{\rm A}}\over{M}}
(\ve{k}\ve{R})\ \ve{m}_{\rm B}^{\rm orb} (0) \ - \ 
{{1}\over{2}}\lambda {{M_{\rm A}}\over{M}} \ve{R} \times 
\left[  H_{\rm B} , \ve{d}_{\rm B}^{(1)} (\ve{k})\ \right] 
\ . 
\label{X_27_result}
\end{equation} 

Since transitions between states of the same parity 
cannot be made with the operators 
$\ve{d}_{\rm A,B}(0)$, 
the 4$^{\rm th}$ and 5$^{\rm th}$ terms in 
Eq.~(\ref{MintClust}) of the lowest (first) order in ${q}$, provide 
the contribution 
\begin{equation}
\ve{X}_{45} (\lambda\ve{k}) = 
i \lambda {{M_{\rm B} \ve{d}_{\rm A}^{(1)}(\ve{k}) 
- M_{\rm A} \ve{d}_{\rm B}^{(1)}(\ve{k})}\over{M}} 
\times \ve{V}\ .  
\label{X_45_result}
\end{equation} 

Substitution of Eqs.~(\ref{X_16_result})~--  
(\ref{X_45_result}) and the same-order contribution given by 
Eqs.~(\ref{M_orb_AB}) into Eq.~(\ref{GMDM}) and taking into account 
the spin contribution coming from Eq.~(\ref{MspinClust}) gives 
\begin{eqnarray}
&&
-i \ve{M}^{M2} (\ve{k})= 
-i (\ve{k}\ve{R})\ \left( 
{{M_{\rm B}}\over{M}} \vegr{\mu}^{\rm A} - {{M_{\rm A}}\over{M}}
\vegr{\mu}^{\rm B}  \right) 
\nonumber\\
&&
+\ {{1}\over{6}}\ve{R}\times\left( 
{{M_{\rm B}}\over{M}}\left[ H_{\rm A} , \ve{d}_{\rm A}^{(1)} (\ve{k})\ \right] - 
{{M_{\rm A}}\over{M}}\left[ H_{\rm B} , \ve{d}_{\rm B}^{(1)} (\ve{k})\ \right]
\right)
\nonumber\\
&&
-\ {{i}\over{3}} 
{{M_{\rm B}\ve{d}_{\rm A}^{(1)} (\ve{k}) - M_{\rm A}\ve{d}_{\rm B}^{(1)} 
(\ve{k}) }\over{M}} \times \ve{V}\ -
i {{e}\over{2m_{\rm N}}} C_{M2} \{ \ve{L} , (\ve{k}\ve{R})\} 
\ ,
\label{M2general_appendix}
\end{eqnarray} 
\begin{equation}
C_{M2} =  {1\over{3 A_{\rm tot} M}} \left( {{M_{\rm B}^2}\over{M_{\rm A}}} Z_{\rm A} - 
{{M_{\rm A}^2}\over{M_{\rm B}}} Z_{\rm B} \right) \ .
\label{C_M2_appendix}
\end{equation} 
%
\section{Appendix: EM transition matrix elements}
\label{App_B}

In this Appendix, we present the detailed formulae for the transition matrix 
elements of the EM operator constructed in Sec.~\ref{sec3} between 
the initial (scattering) and final (bound) states described in Sec.~\ref{sec4} 
for the radiative capture process, Eq.~(\ref{reaction}). 
The notation is as used in those sections. 

Spherical components of the matrix elements of the operators given in
Eqs.~(\ref{D(1)}) and (\ref{M(1)}) between the states specified in
Eqs.~(\ref{Scattstate}) and (\ref{Boundstate}) 
can be written as 
\begin{eqnarray}
&&
\langle f\mid {D}_{\xi}(\ve{k})\mid i \rangle = 
{{E_i^{\rm int} - E_f^{\rm int}}\over {E_i - E_f}} 
\bigg\{ 
eC_{E1} \left[{R}_{\xi}\right]_{if} - 
{{i}\over{6}} (-1)^{\kappa} k_{-\kappa} \left( 
\left[ t_{\xi\kappa}^{\rm A}\right]_{if} + 
\left[ t_{\xi\kappa}^{\rm B}\right]_{if} 
\right.
\nonumber 
\\ 
&&\hspace*{6.0cm}
\left. 
+ 
e C_{E2} \left[ T_{\xi\kappa}^{\rm AB}\right]_{if} 
\right)
\bigg\} \ , 
\label{Dif_spher}
\end{eqnarray} 
\begin{eqnarray}
&&
\langle f\mid {M}_{\xi}(\ve{k})\mid i \rangle = 
\left[ {\mu}_{\xi}^{\rm A}\right]_{if} + 
\left[ {\mu}_{\xi}^{\rm B}\right]_{if} + 
{\mu_{\rm N}} C_{M1} \left[ {L}_{\xi}\right]_{if} 
\nonumber 
\\ 
&&
- i (-1)^{\kappa} k_{-\kappa} \left( 
{{M_{B}}\over{M}}\left[ {\mu}_{\xi}^{\rm A} R_{\kappa}\right]_{if} - 
{{M_{A}}\over{M}}\left[ {\mu}_{\xi}^{\rm B} R_{\kappa}\right]_{if} + 
{\mu_{\rm N}}  C_{M2} \left[\{ {L}_{\xi} , {R}_{\kappa} \}\right]_{if} 
\right)
\ , 
\label{Mif_spher}
\end{eqnarray} 
\begin{displaymath}
( \xi ,\kappa = -1, 0, 1 )
\end{displaymath}
where, for simplicity, we ignore the last two terms of 
expression~(\ref{M2general}), which either do not contribute 
at all, or are negligibly small for the energy range covered by our 
calculation of this amplitude presented in Sec.~\ref{sec5}. 
Then, the matrix elements 
\begin{equation}
\left[ {\cal O} \right]_{if} \equiv 
\langle \Phi^{\rm C}_{J_{\rm C}} J_{\rm C}m_{\rm C} \mid 
{\cal O} 
\mid \Psi^{(+)}_{\ve{P}};
\phi_{J_{\rm A}}^{\rm A} J_{\rm A}m_{\rm A};
\phi_{J_{\rm B}}^{\rm B} J_{\rm B}m_{\rm B}\rangle 
\ , 
\label{Oif_def1}
\end{equation} 
which enter Eqs.~(\ref{Dif_spher}) and (\ref{Mif_spher}), can be written 
as 
\begin{equation}
\left[ {\cal O} \right]_{if} =
{{\omega_{L}}\over{\sqrt{4\pi }}} \sum_{a,a',a'',J} 
C(L 0 J_{\rm A} m_{\rm A} J'''_{\rm A} m_{\rm A} J_{\rm B} m_{\rm B} J , 
m_{\rm A} + m_{\rm B} )\ 
\widehat{J}\ \widehat{L}\ 
\left[ {\cal O} \right]_{a'a''a}^{J J_{\rm C}m_{\rm C}} 
\ , 
\label{Oif_def2}
\end{equation} 
where  
$$
\widehat{j} \equiv \sqrt{2j+1} 
$$
and 
$$
{\omega_{L}} = \left\{ 
\begin{array}{ll}
1 & \textrm{if A is projectile, and B is target} \\
(-1)^{L} & \textrm{otherwise}\ .
\end{array}
\right.
$$
Here, the angular-momentum quantization axis is 
taken to be in the direction of the momentum $\ve{P}$. 
Denoting the radial overlap integral as 
\begin{equation}
I_{a' a'' a}^{J_{\rm C} J (\nu )} \equiv 
\int \Phi_{J_{\rm C} a'}^{\rm C}(R)\ \Psi_{a'' a}^{J\ (+)}(R) \ 
R^{2+\nu }\ {\rm d} R \ ,
\label{overlap_def}
\end{equation}
for the angular-momentum coupling scheme defined by Eq.~(\ref{channelnotation}), 
we have
\begin{eqnarray}
&&
\left[ {R}_{\xi} \right]_{a'a''a}^{J J_{\rm C}m_{\rm C}} = 
\delta_{J'_{\rm A}J''_{\rm A}} \delta_{J'_{\rm B} J''_{\rm B}} 
(-1)^{ J + J' +J'' +L' +J'_{\rm A} +J'_{\rm B}}\ 
\widehat{J'}\ \widehat{J''}\ \widehat{L''} 
\ \langle L'' 0 1 0 \mid L' 0 \rangle 
\nonumber
\\
&&
\times 
\langle J, m_{\rm A}+m_{\rm B}, 1, \xi \mid J_{\rm C}m_{\rm C} \rangle 
\left\{
\begin{array}{ccc}
L'' & J'_{\rm A} & J''\\
J' & 1 & L'
\end{array}
\right\}
\left\{
\begin{array}{ccc}
J'' & J'_{\rm B} & J\\
J_{\rm C} & 1 & J'
\end{array}
\right\}
I_{a' a'' a}^{J_{\rm C} J (1)}
\ , 
\label{[R]}
\end{eqnarray} 
\begin{eqnarray}
&&
\left[ t_{\xi\kappa}^{\rm A} \right]_{a'a''a}^{J J_{\rm C}m_{\rm C}} = 
\sqrt{{3}\over{2}} \delta_{L' L''} \delta_{J'_{\rm B} J''_{\rm B}} 
(-1)^{ J -L' +2J'_{\rm A} +J''_{\rm A} +J'_{\rm B}}\ 
\widehat{J'}\ \widehat{J''}\ 
\ \langle 1 \xi 1 \kappa \mid 2, \xi + \kappa \rangle 
\nonumber
\\
&&
\times 
\langle J, m_{\rm A}+m_{\rm B}, 2, \xi +\kappa \mid J_{\rm C}m_{\rm C} \rangle 
\left\{
\begin{array}{ccc}
J''_{\rm A} & L' & J''       \\
J'          & 2  & J'_{\rm A}
\end{array}
\right\}
\left\{
\begin{array}{ccc}
J'' & J'_{\rm B} & J\\
J_{\rm C} & 2 & J'
\end{array}
\right\}
\nonumber
\\
&&
\times 
\langle \phi_{J'_{\rm A}}^{\rm A} \mid\mid {\cal Q}_{\rm A}
\mid\mid \phi_{J''_{\rm A}}^{\rm A}\rangle\ 
I_{a' a'' a}^{J_{\rm C} J (0)}
\ , 
\label{[t_A]}
\end{eqnarray} 
\begin{eqnarray}
&&
\left[ t_{\xi\kappa}^{\rm B} \right]_{a'a''a}^{J J_{\rm C}m_{\rm C}} = 
\sqrt{{3}\over{2}} \delta_{J' J''} \delta_{L' L''} \delta_{J'_{\rm A} J''_{\rm A}} 
(-1)^{ 2J'_{\rm B} +J''_{\rm B} -J' -J_{\rm C}}\ 
\ \langle 1 \xi 1 \kappa \mid 2, \xi + \kappa \rangle 
\nonumber
\\
&&
\times 
\langle J, m_{\rm A}+m_{\rm B}, 2, \xi +\kappa \mid J_{\rm C}m_{\rm C} \rangle 
\left\{
\begin{array}{ccc}
J''_{\rm B} & J' & J       \\
J_{\rm C}   & 2  & J'_{\rm B}
\end{array}
\right\}
\langle \phi_{J'_{\rm B}}^{\rm B} \mid\mid {\cal Q}_{\rm B}
\mid\mid \phi_{J''_{\rm B}}^{\rm B}\rangle\ 
I_{a' a'' a}^{J_{\rm C} J (0)}
\ , 
\label{[t_B]}
\end{eqnarray} 
\begin{eqnarray}
&&
\left[ T_{\xi\kappa}^{\rm AB} \right]_{a'a''a}^{J J_{\rm C}m_{\rm C}} = 
\sqrt{6}\delta_{J'_{\rm A}J''_{\rm A}} \delta_{J'_{\rm B} J''_{\rm B}} 
(-1)^{ J +J' +J'' +L' +J'_{\rm A} +J'_{\rm B}}\ 
\widehat{J'}\ \widehat{J''}\ \widehat{L''} 
\ \langle 1 \xi 1 \kappa \mid 2, \xi + \kappa \rangle 
\nonumber
\\
&&
\times 
\langle J, m_{\rm A}+m_{\rm B}, 2, \xi + \kappa \mid J_{\rm C}m_{\rm C} \rangle 
\left\{
\begin{array}{ccc}
L'' & J'_{\rm A} & J''\\
J' & 2 & L'
\end{array}
\right\}
\left\{
\begin{array}{ccc}
J'' & J'_{\rm B} & J\\
J_{\rm C} & 2 & J'
\end{array}
\right\}
\nonumber
\\
&&
\times 
\langle L'' 0 2 0 \mid L' 0 \rangle \ 
I_{a' a'' a}^{J_{\rm C} J (2)}
\ , 
\label{[T_AB]}
\end{eqnarray} 
\begin{eqnarray}
&&
\left[ {\mu}_{\xi}^{\rm A} \right]_{a'a''a}^{J J_{\rm C}m_{\rm C}} = 
\delta_{L' L''} \delta_{J'_{\rm B} J''_{\rm B}} 
(-1)^{ J -L' +2J'_{\rm A} +J''_{\rm A} +J'_{\rm B}}\ 
\widehat{J'}\ \widehat{J''}\ 
\langle J, m_{\rm A}+m_{\rm B}, 1, \xi  \mid J_{\rm C}m_{\rm C} \rangle 
\nonumber
\\
&&
\times 
\left\{
\begin{array}{ccc}
J''_{\rm A} & L' & J''       \\
J'          & 1  & J'_{\rm A}
\end{array}
\right\}
\left\{
\begin{array}{ccc}
J'' & J'_{\rm B} & J\\
J_{\rm C} & 1 & J'
\end{array}
\right\}
\langle \phi_{J'_{\rm A}}^{\rm A} \mid\mid {\cal M}_{\rm A}
\mid\mid \phi_{J''_{\rm A}}^{\rm A}\rangle\ 
I_{a' a'' a}^{J_{\rm C} J (0)}
\ , 
\label{[mu_A]}
\end{eqnarray} 
\begin{eqnarray}
&&
\left[ {\mu}_{\xi}^{\rm B} \right]_{a'a''a}^{J J_{\rm C}m_{\rm C}} = 
\delta_{J' J''} \delta_{L' L''} \delta_{J'_{\rm A} J''_{\rm A}} 
(-1)^{ 2J'_{\rm B} +J''_{\rm B} -J' -J_{\rm C} -1 }\ 
\langle J, m_{\rm A}+m_{\rm B}, 1, \xi  \mid J_{\rm C}m_{\rm C} \rangle 
\nonumber
\\
&&
\times 
\left\{
\begin{array}{ccc}
J''_{\rm B} & J' & J       \\
J_{\rm C}   & 1  & J'_{\rm B}
\end{array}
\right\}
\langle \phi_{J'_{\rm B}}^{\rm B} \mid\mid {\cal M}_{\rm B}
\mid\mid \phi_{J''_{\rm B}}^{\rm B}\rangle\ 
I_{a' a'' a}^{J_{\rm C} J (0)}
\ , 
\label{[mu_B]}
\end{eqnarray} 
\begin{eqnarray}
&&
\left[ {L}_{\xi} \right]_{a'a''a}^{J J_{\rm C}m_{\rm C}} = 
\delta_{J'_{\rm A}J''_{\rm A}} \delta_{J'_{\rm B} J''_{\rm B}} \delta_{L' L''} 
(-1)^{ J + J' +J'' +L' +J'_{\rm A} +J'_{\rm B}}\ 
\widehat{J'}\ \widehat{J''}\ \widehat{L'} 
\ \sqrt{ L' (L' + 1 )} 
\nonumber
\\
&&
\times 
\langle J, m_{\rm A}+m_{\rm B}, 1, \xi \mid J_{\rm C}m_{\rm C} \rangle 
\left\{
\begin{array}{ccc}
L'' & J'_{\rm A} & J''\\
J' & 1 & L'
\end{array}
\right\}
\left\{
\begin{array}{ccc}
J'' & J'_{\rm B} & J\\
J_{\rm C} & 1 & J'
\end{array}
\right\}
I_{a' a'' a}^{J_{\rm C} J (0)}
\ , 
\label{[L]}
\end{eqnarray} 
\begin{eqnarray}
&&
\left[ {\mu}_{\xi}^{\rm A} R_{\kappa} \right]_{a'a''a}^{J J_{\rm C}m_{\rm C}} = 
\delta_{J'_{\rm B} J''_{\rm B}} 
(-1)^{ J + J' +J'_{\rm B}}\ 
\widehat{J'}\ \widehat{J''}\ \widehat{L''}\ 
\ \langle L'' 0 1 0 \mid L' 0 \rangle 
\nonumber
\\
&&
\times 
\langle \phi_{J'_{\rm A}}^{\rm A} \mid\mid {\cal M}_{\rm A}
\mid\mid \phi_{J''_{\rm A}}^{\rm A}\rangle\ 
I_{a' a'' a}^{J_{\rm C} J (1)} 
\sum_{g} \widehat{g} \ \langle 1 \xi 1 \kappa \mid g, \xi + \kappa \rangle 
\nonumber
\\
&&
\times 
\langle J, m_{\rm A}+m_{\rm B}, g, \xi +\kappa \mid J_{\rm C}m_{\rm C} \rangle 
\left\{
\begin{array}{ccc}
J'' & J'_{\rm B} & J\\
J_{\rm C} & g & J'
\end{array}
\right\}
\left\{
\begin{array}{ccc}
1   & 1            & g   \\
L'  & J'_{\rm A}   & J'  \\
L'' & J''_{\rm A}  & J''
\end{array}
\right\}
\ , 
\label{[mu_A/R]}
\end{eqnarray} 
\begin{eqnarray}
&&
\left[ {\mu}_{\xi}^{\rm B} R_{\kappa} \right]_{a'a''a}^{J J_{\rm C}m_{\rm C}} = 
\delta_{J'_{\rm A} J''_{\rm A}} 
(-1)^{ J'' + L' +J'_{\rm A}}\ 
\widehat{J'}\ \widehat{J''}\ \widehat{L''}\ 
\ \langle L'' 0 1 0 \mid L' 0 \rangle 
\nonumber
\\
&&
\times 
\langle \phi_{J'_{\rm B}}^{\rm B} \mid\mid {\cal M}_{\rm B}
\mid\mid \phi_{J''_{\rm B}}^{\rm B}\rangle\ 
\left\{
\begin{array}{ccc}
L'' & J'_{\rm A} & J''\\
J'  & 1 & L'
\end{array}
\right\}
I_{a' a'' a}^{J_{\rm C} J (1)} 
\nonumber
\\
&&
\times 
\sum_{g} (-1)^{g} \widehat{g} \ \langle 1 \xi 1 \kappa \mid g, \xi + \kappa \rangle 
\langle J, m_{\rm A}+m_{\rm B}, g, \xi +\kappa \mid J_{\rm C}m_{\rm C} \rangle 
\!\left\{
\begin{array}{ccc}
1   & 1            & g         \\
J'  & J'_{\rm B}   & J_{\rm C} \\
J'' & J''_{\rm B}  & J 
\end{array}
\right\}
\ \!\!, 
\label{[mu_B/R]}
\end{eqnarray} 
\begin{eqnarray}
&&
\left[ \{ {L}_{\xi} , {R}_{\kappa} \} \right]_{a'a''a}^{J J_{\rm C}m_{\rm C}} = 
\delta_{J'_{\rm A}J''_{\rm A}} \delta_{J'_{\rm B} J''_{\rm B}} 
(-1)^{ J + J' +J'' +L'' +J'_{\rm A} +J'_{\rm B}} 
\widehat{J'} \widehat{J''} \widehat{L''} 
\langle L'' 0 1 0 \mid L' 0 \rangle 
I_{a' a'' a}^{J_{\rm C} J (1)}
\nonumber
\\
&&
\times 
\sum_{g} \widehat{g} \langle 1 \xi 1 \kappa \mid g, \xi + \kappa \rangle 
\langle J, m_{\rm A}+m_{\rm B}, g, \xi +\kappa\!\! 
\mid \!\! J_{\rm C}m_{\rm C} \rangle 
\left\{
\begin{array}{ccc}
L'' &  J'_{\rm A} &  J''\\
J'  &  g &  L'
\end{array}
\right\}
\left\{
\begin{array}{ccc}
J'' & J'_{\rm B} & J\\
J_{\rm C} & g & J'
\end{array}
\right\}
\nonumber
\\
&&
\times 
\left( 
(-1)^{g} \widehat{L'}\ \sqrt{ L' ( L' + 1 )} 
\left\{
\begin{array}{ccc}
1  & 1   & g \\
L' & L'' & L'
\end{array}
\right\}
+
\widehat{L''}\ \sqrt{ L'' ( L'' + 1 )} 
\left\{
\begin{array}{ccc}
1  & 1   & g \\
L' & L'' & L''
\end{array}
\right\}
\right) 
\ , 
\label{[LR+RL]}
\end{eqnarray} 
where quantities $\langle \phi_{J'_{\rm A,B}}^{\rm A,B} \mid\mid {\cal M}_{\rm A,B}
\mid\mid \phi_{J''_{\rm A,B}}^{\rm A,B}\rangle$ and 
$\langle \phi_{J'_{\rm A,B}}^{\rm A,B} \mid\mid {\cal Q}_{\rm A,B} 
\mid\mid \phi_{J''_{\rm A,B}}^{\rm A,B}\rangle$ are the reduced matrix elements 
of the magnetic dipole and electric quadrupole operators for cluster A, B. 
In a particular case of transitions that do not change the fragment 
intrinsic 
states ($\phi_{J'_{\rm A,B}}^{\rm A,B} = \phi_{J''_{\rm A,B}}^{\rm A,B}$), 
they are related to the static magnetic dipole ($\mu_{\rm A,B}$) and 
electric quadrupole ($Q_{\rm A,B}$) moments of the nucleus A, B through 
the corresponding definitions~\cite{BrussGlaud} 
\begin{eqnarray}
\frac{\langle J'_{\rm A,B} J'_{\rm A,B}\ 1\ 0 \mid J'_{\rm A,B} J'_{\rm A,B} \rangle}
{\widehat{J'}_{\rm A,B}} \ 
\langle \phi_{J'_{\rm A,B}}^{\rm A,B} \mid\mid {\cal M}_{\rm A,B}
\mid\mid \phi_{J'_{\rm A,B}}^{\rm A,B}\rangle &=& 
\mu_{\rm N}\mu_{\rm A,B}
\nonumber\\
\frac
{\langle J'_{\rm A,B} J'_{\rm A,B}\ 2\ 0 \mid J'_{\rm A,B} J'_{\rm A,B} \rangle}
{\widehat{J'}_{\rm A,B}} \ 
\langle \phi_{J'_{\rm A,B}}^{\rm A,B} \mid\mid {\cal Q}_{\rm A,B}
\mid\mid \phi_{J'_{\rm A,B}}^{\rm A,B}\rangle &=& 
e Q_{\rm A,B} \ .
\label{red_Q}
\end{eqnarray}


\end{document}